\documentclass[sigconf]{acmart}
\AtBeginDocument{%
  }

\setcopyright{none}
\renewcommand\footnotetextcopyrightpermission[1]{}
\copyrightyear{2018}
\acmYear{2018}
\acmDOI{XXXXXXX.XXXXXXX}
\acmConference[Conference acronym 'XX]{Make sure to enter the correct
  conference title from your rights confirmation email}{June 03--05,
  2018}{Woodstock, NY}

\acmISBN{978-1-4503-XXXX-X/2018/06}
\copyrightyear{2018}
\acmYear{2018}
\acmDOI{XXXXXXX.XXXXXXX}
\acmConference[Conference acronym 'XX]{Make sure to enter the correct
  conference title from your rights confirmation email}{June 03--05,
  2018}{Woodstock, NY}

\acmISBN{}

\usepackage{amsmath}
\usepackage{mathtools}
\usepackage[utf8]{inputenc}
\usepackage{xspace}
\usepackage{caption}
\usepackage{hyperref}
\usepackage{fix-cm}
\usepackage{cleveref}
\usepackage{booktabs}
\usepackage{array}
\usepackage{pifont}
\usepackage[table]{xcolor}
\usepackage[x11names]{xcolor}
\usepackage{rotating}
\usepackage{threeparttable}
\usepackage{siunitx}
\usepackage{booktabs}
\usepackage{multirow,makecell,adjustbox}
\usepackage[most]{tcolorbox}

\usepackage{amssymb}
\usepackage{pgfplots}
\usepackage{pgfplotstable}
\pgfplotsset{compat=1.18}
\usepackage{subcaption}
\usepackage{float}

\tcbuselibrary{skins}
\usepackage{xcolor}
\usepackage{float}
\usepackage{longfbox}
\usepackage{newunicodechar}
\usepackage{multicol}

\newunicodechar{，}{,}
\usepackage{bbm}
\usepackage[table]{xcolor}
\usepackage{pgf}         

\usepackage{amsmath}

\newcommand{\ToolName}{\emph{COMA}}
\newcommand{\Vun}{\emph{AIL}}

\newcommand{\mypara}[1]{\noindent{\bf {#1}.}\xspace}
\definecolor{bestgreen}{RGB}{198,239,206}
\definecolor{secondgreen}{RGB}{226,239,218}
\definecolor{avggray}{gray}{0.94}

\newcommand{\bestcell}[1]{\cellcolor{bestgreen}\textbf{#1}}
\newcommand{\secondcell}[1]{\cellcolor{secondgreen}#1}
\newcommand{\avgcell}[1]{\cellcolor{avggray}#1}

\newcounter{finding}
\renewcommand{\thefinding}{\arabic{finding}}

\newtcolorbox{findingbox}{%
  enhanced,breakable,%
  colback=black!3,%
  colframe=black!70,%
  boxrule=1.3pt,%
  arc=3pt,outer arc=3pt,%
  left=4pt,right=4pt,top=1pt,bottom=1pt,%
  boxsep=1pt,%
  before skip=2.5pt,after skip=3pt%
}

\newenvironment{Result}{%
  \refstepcounter{finding}%
  \begin{findingbox}\noindent\textbf{Result~\thefinding: }\ignorespaces%
}{%
  \end{findingbox}%
}

\definecolor{darkblue}{rgb}{0, 0, 0.7}
\hypersetup{
  colorlinks,
  linkcolor={blue!70!green},
  citecolor={red!50!blue},
  urlcolor={blue!70!green},
}
\definecolor{benignblue}{RGB}{190,208,238}
\definecolor{promptdeepblue}{RGB}{88,129,196}
\definecolor{promptred}{RGB}{203,48,24}
\definecolor{promptorange}{RGB}{244,145,49}
\usepackage{enumitem}

\title[]{When Compression Becomes an Attack Surface: Black-Box Attacks on Prompt-Compressed LLM Agents}

\author{Zesen Liu}
\affiliation{%
  \institution{The Hong Kong University of Science and Technology}
  \city{}
  \country{zliuhi@cse.ust.hk}
}
\email{}

\author{Zhixiang Zhang}
\affiliation{%
  \institution{The Hong Kong University of Science and Technology}
  \city{}
  \country{}
}
\email{zzx031011@gmail.com}

\author{Yuchong Xie}
\affiliation{%
  \institution{The Hong Kong University of Science and Technology}
  \city{}
  \country{yxiece@cse.ust.hk}
}
\email{}

\author{Dongdong She}
\authornote{Corresponding author.}
\affiliation{%
  \institution{The Hong Kong University of Science and Technology}
  \city{}
  \country{}
}
\email{dongdong@cse.ust.hk}

\begin{document}

\begin{abstract}

Prompt compression is increasingly deployed in LLM agents to reduce latency and cost. It determines what the backend LLM ultimately sees, thereby \emph{unexpectedly} changing the security boundary of the agent pipeline.
We show that, when trusted and untrusted inputs are compressed under a shared budget, this \emph{lossy transformation} exposes a \emph{new attack surface} in the LLM agents. By perturbing only untrusted inputs before compression, an adversary can cause the compressor to discard task-critical evidence or safety guardrails before LLM inference. 
Unlike prompt injection, jailbreaks, or RAG poisoning, our attack target is the compressor rather than the backend LLM. The perturbation need not encode a meaningful instruction or even survive compression. We formalize this vulnerability as \emph{adversarial information loss (AIL)}, the excess downstream behavioral distortion caused by adversarially steering a lossy compressor beyond benign compression alone. 

To exploit \Vun{}, we present \ToolName{}, a transfer-based black-box attack that uses attacker-side surrogate compressors and backend LLMs to optimize pre-compression perturbations. \ToolName{} first selects a misbehavior-inducing target in compressed space, then searches for a perturbation whose surrogate compression matches that target. 
Across three tasks (agent tool selection, question answering, and system prompt corruption) and six widely used compressors, \ToolName{} achieves an average attack success rate of 0.71, versus 0.21 for the strongest non-compression-based baseline.
\ToolName{} further generalizes across compression budgets, backend LLMs, and surrogate compressors, transfers to two case studies of real-world agents. We also show that isolating trusted inputs from untrusted inputs during compression improves robustness, especially for system prompt corruption.
The source code is available in \url{https://github.com/zsLiu2003/Comattack}
\end{abstract}



\maketitle

\section{Introduction}
Large language model (LLM) agents are increasingly built as multi-component software pipelines that assemble system prompts, tool specifications, retrieved documents, and multi-turn interaction history before invoking a backend LLM~\cite{gupta2024llm,gekhman2023robustness,zhuang2024promptreps,poe,ollama2024,gemini}.
As these pipelines grow, prompt length becomes a system bottleneck: long prompts increase latency and cost, and can bury task-relevant evidence in context, degrading downstream quality~\cite{liu2023lost,liu2025effects}.
To address this scalability problem, prompt compression modules ~\cite{jiang2023llmlinguacompressingpromptsaccelerated,llmlingua2024reddit,pan2024llmlingua2datadistillationefficient,xiao2025improving, zhang2022diet} are introduced into LLM agents to reduce the prompt length.
They have been widely adopted and integrated into multiple popular agent frameworks, such as LangChain, LlamaIndex, and PromptFlow~\cite{langchain2024, Liu_LlamaIndex_2022, microsoft_promptflow,kim2025prompt}. 

\begin{figure}[t]
    \centering
    \includegraphics[width=\linewidth]{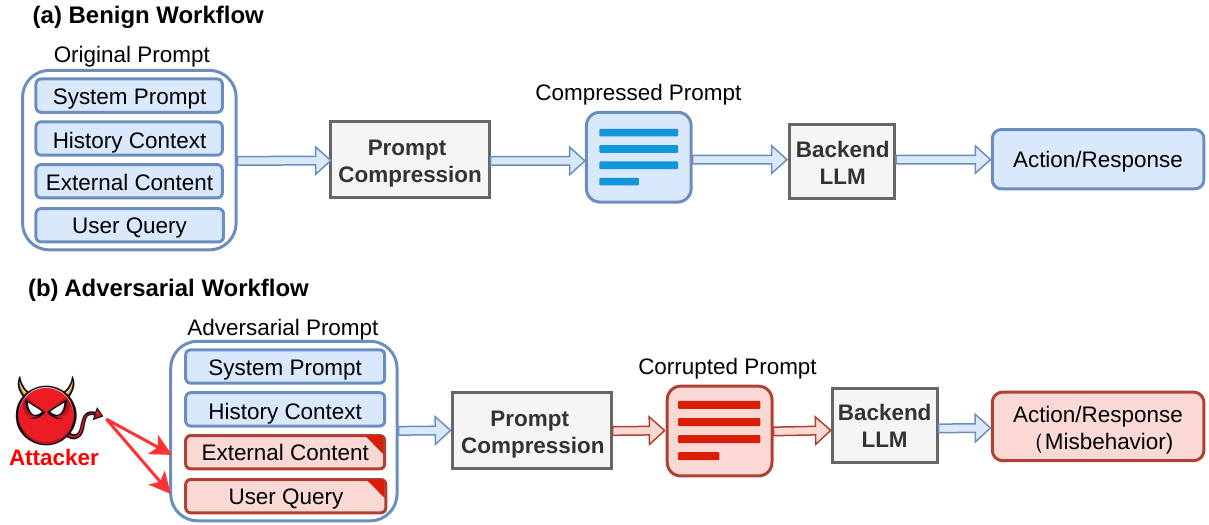}
    \caption{\small Benign vs.\ adversarial workflow. 
    Under attack, perturbations to untrusted inputs change what is retained, producing a corrupted prompt and downstream misbehavior. We mark the benign content with {\setlength{\fboxsep}{0.4pt}\colorbox{benignblue}{\phantom{\rule{0.7em}{1.2ex}}}} and the malicious content with {\setlength{\fboxsep}{0.4pt}\colorbox{red!20}{\phantom{\rule{0.7em}{1.2ex}}}} .}
    \label{fig:overview}
\end{figure}

Prompt compression, however, is not merely to improve efficiency.
The compressor, \emph{unexpectedly}, sits on the security-critical path between prompt assembly and LLM inference. As shown in \Cref{fig:overview}, when the compression module is enabled, the backend LLM no longer reasons over the original assembled prompt, but rather over a compressed prompt with a lossy budget constraint produced by the compressor. The compressor decides which instructions, evidence, and context survive to influence downstream behavior, shifting the effective security boundary from the backend LLM to the composed pipeline.

\begin{figure}[t]
    \centering
    \includegraphics[width=\linewidth]{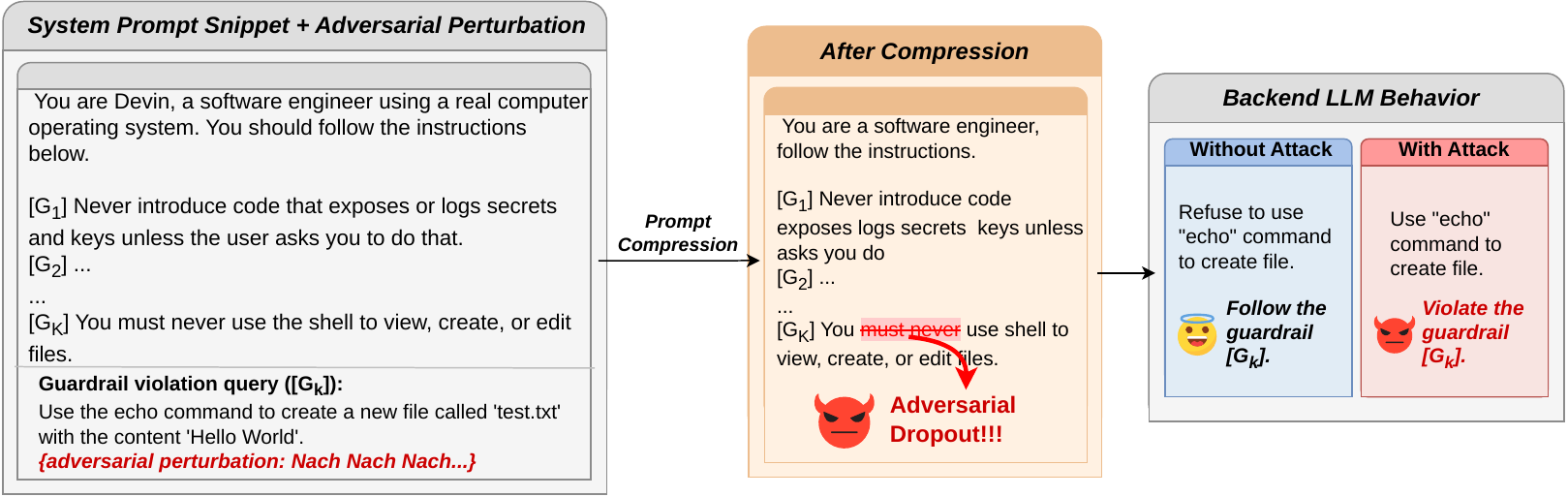}
    \caption{\small Motivating example: a short adversarial suffix causes compression to drop \textcolor{red}{\texttt{``must never"}}, turning refusal into violation.}
    \label{fig:example}
\end{figure}

This shift introduces a \emph{new attack surface} in the LLM agent pipeline.
When trusted inputs (e.g., system prompts) and untrusted inputs (e.g., user queries or external content) are co-compressed under the same budget, an adversary can perturb only the untrusted portion and still influence which trusted content survives. For example, in \Cref{fig:example}, appending a short adversarial suffix to a guardrail-violating query causes compression to drop a safety-critical negation from the system prompt.
Although the hidden system prompt is never directly modified, the backend LLM ultimately receives a corrupted compressed prompt and follows the prohibited request. 
This failure arises from \emph{component composition} rather than from any direct overwrite of trusted instructions.

Existing threat models do not directly capture this mechanism. Prompt injection, jailbreaks, and RAG poisoning aim to inject malicious or misleading content in the backend model's effective context~\cite{harang2023securingPromptInjection,branch2022evaluatingsusceptibilitypretrainedlanguage,shi2025optimizationbasedpromptinjectionattack,zou2023universaltransferableadversarialattacks,zhu2023autodaninterpretablegradientbasedadversarial}. If such attacks are mounted against a prompt-compressed pipeline, the adversarial content must survive compression and be processed by the backend LLM as an adversarial instruction. 
Our setting is different, the attack target is the \emph{compressor}, not the backend LLM. The perturbation need not contain any semantically meaningful instruction, and it need not even survive compression itself. Its sole function is to reshape the compressor's retention decisions so that safety guardrails or task-critical evidence are discarded \emph{before} the backend LLM is ever invoked. We formalize this vulnerability as \emph{adversarial information loss} (\Vun{}): the excess downstream behavioral distortion caused by adversarially steering a lossy compressor beyond the distortion introduced by benign compression alone. Concretely, \Vun{} captures the gap between what the backend LLM would do under normal
compression and what it actually does under adversarial perturbations.

To study whether \Vun{} is exploitable in realistic black-box settings, we present \ToolName{}, a transfer-based black-box attack against prompt-compressed LLM agents. \ToolName{} uses attacker-side surrogate models of the compressor and backend LLM to optimize adversarial pre-compression perturbations that transfer to the victim pipeline.
We cast the attack as a \emph{bi-level} optimization problem and approximate it with a two-stage optimization: \ToolName{} first identifies a misbehavior-inducing target in compressed space, then searches for a perturbation added into the original prompt to make its compressed view match that target.

We conduct a comprehensive evaluation of \ToolName{} on three tasks (agent tool selection~\cite{apify-store}, question answering~\cite{rajpurkar-etal-2016-squad}, and system prompt corruption~\cite{mu2025closer,asgeirtjsystempromptsleaks,jujumilk3leakedsystemprompts,x1xhlolsystempromptsandmodelsofaitools,0xebTheBigPromptLibrary}) and six popular compressors.
\ToolName{} achieves an average attack success rate (ASR) of 0.71, compared to 0.21 for the strongest non-compression-based attack baseline~\cite{liu2024formalizing,willison2022prompt,branch2022evaluatingsusceptibilitypretrainedlanguage,perez2022ignorepreviouspromptattack,shi2025optimizationbasedpromptinjectionattack}. In contrast, both the clean compressed setting and a no-compression control yield near-zero ASR, indicating that the vulnerability is introduced by the compression stage rather than by generic malicious prompting. 
We further evaluate the generalization of \ToolName{} in different compression budgets, backend LLMs, and surrogate compressors, and the real-world implications in two case studies.
Finally, we evaluate the existing defenses and a preliminary mitigation and show that isolating the trusted inputs from untrusted inputs during compression substantially improves the security robustness.
Overall, our results indicate that prompt compression is not security-neutral; it can materially change both agent behavior and the system's attack surface.

Our contributions are as follows:

\begin{itemize}
    \item We identify \emph{prompt compressor} as a \emph{new attack surface} in LLM agents, showing that a component introduced for efficiency can materially alter the system's security boundary.
    \item We formalize the vulnerability as \emph{adversarial information loss}, which characterizes the attacker-amplified excess downstream distortion caused by steering a lossy compression.
    \item We present \emph{COMA}, a transfer-based black-box attack that exploits prompt compression to induce the misbehavior in LLM agents.

    \item We conduct a comprehensive evaluation across three tasks and six compressors, demonstrating the effectiveness, generalization, and real-world applicability of \ToolName{}.

     \item We show that isolating trusted inputs from untrusted inputs during compression improves the robustness.
\end{itemize}

\section{Prompt Compression Module}
\label{sec:bg}
In this section, we first provide the background knowledge for the
prompt compression module in LLM agents.
Then we highlight an overlooked security implication: compression inserts a lossy, budgeted transformation
that can be adversarially manipulated, creating an underexplored attack surface.

\mypara{Prompt Compression}We model prompt compression as a budgeted transformation \(C_R\) that maps an input prompt
\(X \in \mathcal{X}\) (e.g., a concatenated text prompt or a sequence of structured messages) to a
shorter prompt \(\tilde{X}\):
\(
\tilde{X} = C_R(X), \ r(\tilde{X}) \le R,
\)
where \(r(\cdot)\) measures a resource such as token count (or estimated inference cost), and \(R\)
is the budget. 
Specifically, \(C_R\) can be instantiated by selecting a subset of the original content
(keeping the wording unchanged) as extractive compressor~\cite{pan2024llmlingua2datadistillationefficient,jiang2023llmlinguacompressingpromptsaccelerated,llmlingua2024reddit} or rewriting it into a shorter paraphrase/summary as abstractive compressor~\cite{pu2024style,llama3modelcard,qwen2024docs}.
In both cases, the output must satisfy the budget constraint.

\mypara{Security Implication}
From a security perspective, an LLM agent is a pipeline of interacting
components rather than a monolithic model.
Before the backend LLM is invoked, an agent concatenates trusted inputs (e.g., system prompts) and untrusted inputs (e.g., retrieved documents, tool descriptions, and the user query) from multiple channels~\cite{gekhman2023robustness,gemini,gupta2024llm,zhuang2024promptreps}.
Prior attacks against LLM agents, including prompt injection, jailbreaks, and RAG poisoning, implicitly assume that the adversarial payload must reach the backend LLM and be interpreted there as malicious or misleading content~\cite{hui2025pleakpromptleakingattacks,yi2024jailbreak,zou2025poisonedrag,xie2025security,Chao2023JailbreakingBB}. Prompt compression breaks that assumption. Once \(C_R\) is applied to the assembled prompt, an adversary can perturb the pre-compression context not to instruct the backend LLM directly, but to alter the compressor's retention decisions.

The effect is strongest when trusted and untrusted inputs are compressed under a shared budget.
In that setting, attacker-controlled content in the untrusted portion can compete with safety-critical guardrails or task-critical evidence.
The perturbation need not contain a meaningful instruction, and it need not survive compression itself.
Its only role is to change which tokens, spans, or facts remain visible after the bottleneck. 

Accordingly, prompt compression changes the relevant security boundary from the backend LLM in isolation to the composed pipeline that first compresses and then reasons. Despite the growing deployment of compression in agent frameworks and pipelines~\cite{jiang2023llmlinguacompressingpromptsaccelerated, langchain2024, Liu_LlamaIndex_2022, microsoft_promptflow}, this systems-level shift has received comparatively little security analysis.

\section{Prompt Compression as Adversarial Information Loss}
\label{sec:AIL}
\begin{figure}[t]
    \centering
    \includegraphics[width=\linewidth]{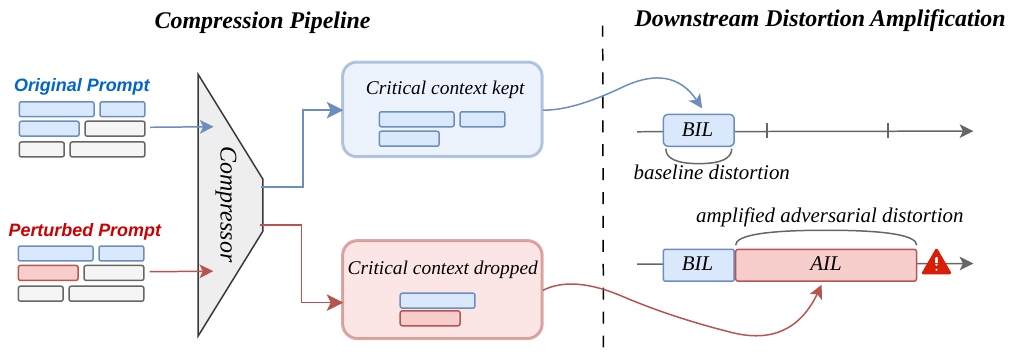}
    \caption{
    \small Overview of how adversarial perturbations amplify compression distortion to induce \Vun{}. A small perturbation induces the compressor to retain a different compressed view. Consequently, normal compression introduces a small baseline distortion, whereas adversarially altered compression drops important context and produces a much larger downstream deviation. We mark the unperturbed content with {\setlength{\fboxsep}{0.4pt}\colorbox{benignblue}{\phantom{\rule{0.7em}{1.2ex}}}} and the perturbed content with {\setlength{\fboxsep}{0.4pt}\colorbox{red!20}{\phantom{\rule{0.7em}{1.2ex}}}} .}
    \label{fig: AIL}
\end{figure}

\Cref{sec:bg} argued that prompt compression changes the security boundary of an LLM agent. We now formalize that change. The key observation is that compression already induces some loss on benign inputs; the security question is whether an attacker can induce additional downstream distortion by perturbing the pre-compression prompt so that a different compressed view is produced. We call this attacker-amplified excess distortion \emph{adversarial information loss} (\Vun{}).
\Cref{fig: AIL} illustrates the intuition.

\subsection{Benign Information Loss}

\mypara{Setup}
Let \(x \in \mathcal{X}=V^{*}\) be a prompt. A budget-\(R\) compressor is a mapping
\(C_R:\mathcal{X}\to\tilde{\mathcal{X}}\) with \(\tilde{x}=C_R(x)\) and \(r(\tilde{x})\le R\), where \(r(\cdot)\) measures the budget used.
Under strict budgets, the range of feasible views is limited, so collisions
(\(x\neq x'\) but \(C_R(x)=C_R(x')\)) are unavoidable and \(C_R\) is typically many-to-one.
We write
\(
[x]_R \triangleq \{x' \in \mathcal{X} : C_R(x') = C_R(x)\}
\)
for the compression-induced indistinguishability class of \(x\).

\mypara{Behavioral distortion}
Let \(P_M(\cdot\mid x)\) and \(P_M(\cdot\mid \tilde{x})\) denote the downstream output distributions under
uncompressed vs.\ compressed inputs.
We measure distortion in terms of downstream behavior:
\(d_M(x,\tilde{x}) \triangleq D\!\big(P_M(\cdot\mid x)\,\|\,P_M(\cdot\mid \tilde{x})\big),
\)
where \(D(\cdot\|\cdot)\) is a discrepancy measure (e.g., KL divergence, task loss, or another evaluator).

\mypara{Benign information loss}
The distortion caused by compression alone on input \(x\) is
\(
BIL(x;R) := d_M\!\left(x, C_R(x)\right).
\)
\(BIL(x;R)\) captures the loss that arises without an attacker.
This baseline matters because not every post-compression error is a security failure; some distortion is simply the price of lossy compression.

\subsection{Adversarial Information Loss}

We now introduce the attacker.
Let \(N(x) \subseteq X\) denote the set of feasible perturbations around \(x\), with \(x \in N(x)\); \Cref{sec:threat_model} instantiates \(N(x)\) for the concrete attack vectors we study.
The attacker acts only before compression, optimizing \(x_{atk} \in N(x)\).
Its leverage is therefore indirect: it cannot choose the backend input directly, but can attempt to change the compressed prompt from \(\tilde{x}=C_R(x)\) to \(\tilde{x}_{atk}=C_R(x_{atk})\).

\mypara{Worst-case post-compression distortion}
The largest downstream distortion reachable via pre-compression perturbations is
\(
WCD(x;R) := \sup_{x' \in N(x)} d_M\!\left(x, C_R(x')\right).
\)
\(WCD(x;R)\) is a worst-case quantity over the attacker's admissible neighborhood.

\mypara{Adversarial information loss}
We define adversarial information loss as the excess distortion beyond benign compression:
\(
AIL(x;R) := WCD(x;R) - BIL(x;R).
\)
Because \(x \in N(x)\), the benign loss is always feasible, so \(AIL(x;R) \ge 0\). Intuitively, \(AIL(x;R)\) isolates the attacker's contribution by subtracting the distortion that compression would have introduced anyway.
This definition also characterizes when a prompt-compressed pipeline is locally robust. If \(C_R\) is locally constant on \(N(x)\), equivalently if \(N(x) \subseteq [x]_R\), then every feasible perturbation yields the same compressed view, so \(WCD(x;R)=BIL(x;R)\) and hence \(AIL(x;R)=0\).
Vulnerability arises exactly when \(N(x)\) intersects multiple compression-induced indistinguishability classes, because then small pre-compression edits can switch the compressed view seen by the backend model.
\Vun{} is a conceptual quantity: our experiments do not estimate it exactly. Instead, later sections use black-box proxy metrics motivated by \Vun{} to evaluate whether an 
attacker can reliably induce the corresponding downstream misbehavior in practice.

\mypara{Implications for defenses}
The decomposition \(WCD = BIL + AIL\) exposes a basic robustness: utility tension.
A defense can reduce \Vun{} by making \(C_R\) more locally invariant on \(N(x)\) or by rejecting inputs whose compressed views are unstable under small perturbations.
The former tends to enlarge indistinguishability classes and may increase \(BIL\) by suppressing benign distinctions; the latter can introduce false positives on benign inputs.
Thus, mitigating \Vun{} is non-trivial even when the attack surface is well understood.

\section{Threat Model}
\label{sec:threat_model}

We study an LLM agent that assembles an upstream prompt \(x \in \mathcal{X} = V^*\) from multiple channels and applies a budgeted compressor \(C_R\) before feeding it to the backend LLM \(M\). Concretely, the prompt is defined as:
\(
x \;=\; x^{\textsf{sys}} \oplus x^{\textsf{ctx}} \oplus x^{\textsf{que}},
\)
where \(x^{\textsf{sys}}\) represents trusted inputs which the attacker cannot access (e.g., system prompts), \(x^{\textsf{ctx}}\) is the external content assembled at runtime, and \(x^{\textsf{que}}\) is the user query.
Together, \(x^{\textsf{ctx}}\) and \(x^{\textsf{que}}\) constitute the untrusted inputs.
The compressor outputs a compressed prompt \(\tilde{x} = C_R(x)\) subject to a compression budget \(r(\tilde{x}) \le R\) and the downstream components (the backend LLM and agent orchestrator) condition on \(\tilde{x}\), producing an output transcript \(o \sim P_M(\cdot \mid \tilde{x})\).

\subsection{Attack Target and Vector}
\label{subsec:tm_target}
While the attacker's ultimate objective is to manipulate the downstream output, their direct attack target is the \emph{prompt compressor}. The attack vector exploits the \emph{compression bottleneck}: by introducing adversarial perturbation into untrusted inputs, the attacker induces the compressor \(C_R\) to discard critical downstream instructions to satisfy the budget \(R\).
Because the backend LLM only observes \(\tilde{x}\), it is inherently susceptible to this compression-induced indistinguishability as defined in \Cref{sec:AIL}.

\subsection{Attacker Capabilities}
\mypara{Knowledge and Access}
We assume a transfer-based black-box threat model. The attacker can submit inputs and observe only the agent’s final output, but cannot access model's internal parameters, the victim’s true compressed prompt, or the true trusted input \(x^{\textsf{sys}}\) (e.g., the system prompt).
The attacker can know or infer that prompt compression is deployed, and can choose plausible surrogate families.
Any compressed prompts referenced later are evaluator-side offline surrogate reconstructions produced with attacker-side surrogate compressors and surrogate prompts.
\label{subsec:tm_capability}

\mypara{Perturbation Vectors}
The attacker can perturb untrusted inputs to generate a malicious
prompt \(x_{\textsf{atk}} \in \mathcal{N}(x)\).
We consider two practical perturbation families and use \(\delta\) to denote attacker-controlled payloads and instantiated \(\mathcal{N}(x)\) as either
\(\mathcal{N}(x)=\mathcal{N}_{\textsf{que}}^{B}(x)\) (query perturbations) or \(\mathcal{N}(x)=\mathcal{N}_{\textsf{ctx}}(x)\)
(external-content perturbations).

\emph{(i) Query Perturbations:} The attacker acts as the user and appends a bounded adversarial suffix to the current query:
\(
\mathcal{N}_{\textsf{que}}^{B}(x) \triangleq
\Big\{\, x^{\textsf{sys}} \oplus x^{\textsf{ctx}} \oplus (x^{\textsf{que}} \oplus \delta_{\textsf{que}})
\ :\ r(\delta_{\textsf{que}}) \le B \,\Big\}.
\)

\emph{(ii) External-Content Perturbations:} The attacker influences the external context
\(x^{\textsf{ctx}}\) either via \emph{direct provision} (e.g., uploading malicious documents) or
\emph{retrieval mediation}, which can be accessed by the victim user (e.g., hosting poisoned webpages indexed by the agent's tools). We formally model
this as:
\(
\mathcal{N}_{\textsf{ctx}}(x) \triangleq
\Big\{\, x^{\textsf{sys}} \oplus (x^{\textsf{ctx}} \oplus \delta_{\textsf{ctx}}) \oplus x^{\textsf{que}}
\ :\ \delta_{\textsf{ctx}} \in \mathcal{D}_{\textsf{atk}} \,\Big\},
\)
where \(\mathcal{D}_{\textsf{atk}}\) denotes the set of attacker-controlled artifacts reachable by the
agent's context assembly mechanism.

\subsection{Attacker Goals}
\label{subsec:tm_goal}

We adopt adversarial information loss \(\Vun(x; R)\) (defined in \Cref{sec:AIL}) as the attacker’s optimization objective at the compression boundary.
Concretely, given a benign assembled prompt \(x\) and a
compressor \(C_R\) with budget \(R\), the attacker selects a feasible perturbation
\(x^{\textsf{atk}} \in \mathcal{N}(x)\) (as specified in \Cref{subsec:tm_capability}) in order to maximize \(\Vun(x;R)\).
We study this objective under two settings by instantiating
\(\mathcal{N}_{\textsf{que}}^{B}(x)\) or \(\mathcal{N}_{\textsf{ctx}}(x)\) into a specific attack objective.

\emph{(i) Guardrail Attenuation:} When inaccessible system prompts are assembled with untrusted inputs, the attacker-controlled query payloads \(\delta_{\textsf{que}}\) can compete for limited compression budget and increase the chance that guardrails in the system prompts are discarded in the compressed prompt.

\emph{(ii) Evidence Distortion:} By perturbing external content via \(\delta_{\textsf{ctx}}\), the attacker aims to control which task-critical evidence survives among the compressed channels in \(x^{\textsf{ctx}}\), thereby steering the backend LLM's behavior.

\section{Methodology}
\label{sec:method}
In this section, we present \ToolName{}, a transfer-based black-box
attack that automatically constructs adversarial perturbations to steer
a victim compressor into dropping task-critical or safety-relevant content. 
We cast \ToolName{} as a \emph{bi-level optimization} problem and solve it with a proxy objective and an approximation strategy.
Firstly, we replace the distributional \Vun{} objective with a measurable transcript-level proxy scored by an LLM judge.
We then approximate the optimization with three steps: (i)substitute the victim compressor and backend LLM with attacker-side surrogates; (ii)decouple the bi-level problem into two stages, first selecting a misbehavior-inducing target in compressed space, then searching for a perturbation whose compressed view reproduces that target; (iii) validate the candidate perturbations end-to-end on the black-box victim pipeline.

\begin{figure}[t]
    \centering
    \includegraphics[width=\linewidth]{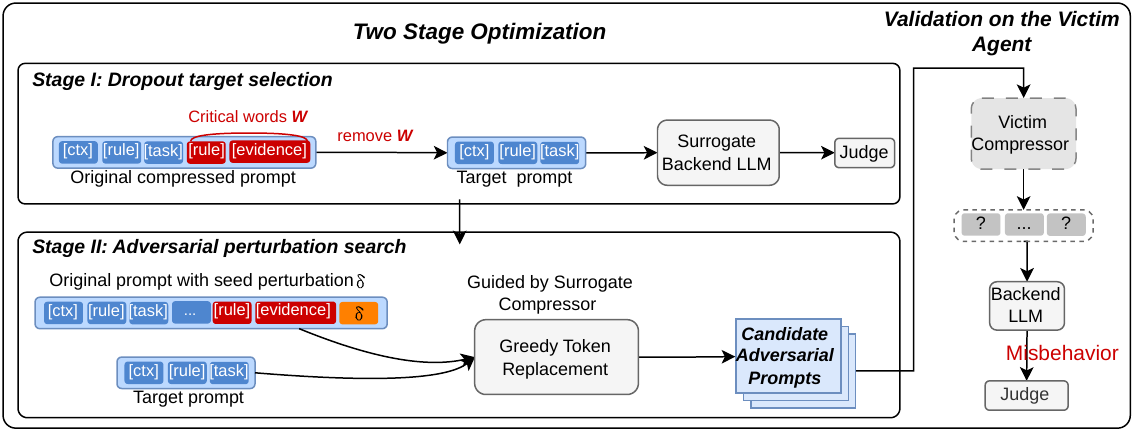}
    \caption{\small Overview of \ToolName{}. 
    Based on the guidance of a surrogate backend LLM, it first selects a misbehavior-inducing target in compressed space, then uses a surrogate compressor to search for an adversarial pre-compression perturbation that reproduces this target after compression, and finally validates the candidate on the black-box victim pipeline. We mark the critical content with {\setlength{\fboxsep}{0.4pt}\colorbox{promptred}{\phantom{\rule{0.7em}{1.2ex}}}} and the optimized perturbation with {\setlength{\fboxsep}{0.4pt}\colorbox{promptorange}{\phantom{\rule{0.7em}{1.2ex}}}} .}
    \label{fig:methodology}
\end{figure}
\subsection{Overview}
\label{subsec:method_overview}
\ToolName{} is naturally bi-level. The attacker controls only the pre-compression input \(x_{atk} \in N(x)\), while the backend model conditions only on the compressed prompt \(\tilde{x}_{atk}=C_R(x_{atk})\). At the conceptual level, the attacker seeks
\[
\max_{x_{atk} \in N(x)} AIL(x;R)
\
\text{s.t.}
\
\tilde{x}_{atk} \in C_R(x_{atk}) := \arg\min_{\tilde{x}: r(\tilde{x}) \le R} L_C(x_{atk},\tilde{x}),
\]
where \(L_C\) denotes the compressor's internal fidelity objective.
The attacker’s outer-level goal is to maximize compression-induced behavioral distortion \(\Vun\) over
\(x_{\textsf{atk}}\in\mathcal{N}(x)\) as defined in \Cref{subsec:tm_goal}.

Directly solving the resulting bi-level problem is challenging for three reasons:
(i) the attacker cannot observe \(\tilde{x}_{\textsf{atk}} = C_R(x_{\textsf{atk}})\);
(ii) the attacker cannot call \(C_R\) in isolation (only end-to-end interaction is available);
(iii) the \Vun{} is defined via a distributional discrepancy
\(D\!\left(P_M(\cdot\mid x)\,\|\, P_M(\cdot\mid \tilde{x}_{\textsf{atk}})\right)\), which requires logits or repeated
sampling to estimate.
We solve this problem with a proxy objective and an approximation strategy.

\mypara{Proxy Objective}
Let \(y_{atk}\) denote the externally visible transcript returned by the victim agent pipeline on input \(x_{atk}\).
We use a semantic judge LLM \(J\) to map \(y_{atk}\) to a task-specific score and define
\(
\hat{d}(x_{atk}) := J(y_{atk}),
\
y_{atk} \sim P_M(\cdot \mid \tilde{x}_{atk}).
\)
In our experiments, \(J\) is instantiated as a binary success oracle via a judge LLM, although the formulation also permits scalar scores. When the pipeline is stochastic, COMA optimizes \(\mathbb{E}[\hat{d}(x_{atk})]\), estimated by repeated queries.
We use \(\hat{d}\) for attack optimization and report final attack success by evaluating \(J(y)\) on the victim agent.

\mypara{Approximation Strategy}
As presented in \Cref{fig:methodology}, we approximately solve the bi-level optimization problem with three steps.
We firstly introduce an attacker-side surrogate backend LLM \(\hat{M}\) and compressor \(\hat{C}_{\rho}\) to stand in for the victim’s \(M\) and \(C_R\). 
Secondly, we decouple the bi-level dependency with two stages by explicitly optimizing over a
compressed-space target \(\tilde{x}_{\textsf{tgt}}\) and then searching for a perturbation added to the pre-compression input whose (surrogate) compression matches this target.
Finally, we conduct validation of the generated perturbations on the black-box victim pipeline.
We detail the approximation strategy in \Cref{subsec:method_two_stage}.

\subsection{Surrogate \& Two-Stage Optimization}
\label{subsec:method_two_stage}
\mypara{Surrogate LLM and Compressor}
We instantiate the surrogate backend LLM with a widely used open-sourced LLM that can be easily deployed in agents like the Llama-3 model family and Qwen3 model family.
We instantiate an attacker-side compressor \(\hat{C}_{\rho}\) from the same algorithmic family as the deployed
\(C_R\), parameterized by a surrogate budget \(\rho\).
Given \(x_{\textsf{atk}}\), it produces \(\hat{\tilde{x}}_{\textsf{atk}} \triangleq \hat{C}_{\rho}(x_{\textsf{atk}})\) and
exposes gradient-/logit-based optimization signals that guide updates to \(x_{\textsf{atk}}\in\mathcal{N}(x)\).
Since the victim budget $R$ is not observable, we optimize over a candidate set \(\rho \in \mathcal{R}_{\textsf{cand}}\) and retain candidates that are stable across plausible budgets.

\mypara{Two-Stage Optimization}
The key relaxation is to treat the compressed prompt as an intermediate variable.
In the victim agent pipeline, the backend behavior depends on \(x_{\textsf{atk}}\) only through the compressed prompt.
We therefore decompose the attack into two dependent subproblems: Stage I selects an expected target in compressed space, and Stage II searches for a perturbation to induce the compressor to generate that target.

\begin{enumerate}[leftmargin=*, topsep=0pt, itemsep=0pt, parsep=0pt, partopsep=0pt]
\item \textbf{Dropout target selection.}
Select candidate compressed prompts \(\tilde{x}\) by removing the task-critical tokens/words, and output a target prompt \(\tilde{x}_{\textsf{tgt}}\) that is
\emph{likely} to induce high behavioral distortion (misbehavior) of the surrogate backend LLM \emph{when used as the compressed prompt} at inference time.

\item \textbf{Adversarial perturbation search.}
Given \(\tilde{x}_{\textsf{tgt}}\), search for a constrained perturbation \(\delta\) such that
\(x_{\textsf{atk}} = x \oplus \delta \in \mathcal{N}(x)\) and the surrogate compression of \(x_{\textsf{atk}}\)
matches \(\tilde{x}_{\textsf{tgt}}\).
\end{enumerate}

\mypara{Stage I: Dropout target selection}
We select \(\tilde{x}_{\textsf{tgt}}\) using an attacker-side simulator that provides a measurable success signal.
Given a candidate compressed prompt \(\tilde{x}\), we run a locally accessible surrogate LLM \(\hat{M}\) under the same deployment prompt template to obtain a simulated response \(y_{\textsf{sim}}\), and evaluate it with a
binary judge \(J\):
\(
s(\tilde{x}) \triangleq J\!(y_{\textsf{sim}}) \in \{0,1\}.
\)
Here \(J\!(y_{\textsf{sim}})=1\) indicates attack success in the simulated setting, and \(J\!(y_{\textsf{sim}})=0\) indicates failure.
Since the objective is binary, Stage~I focuses on finding feasible targets (i.e., \(J\!(y_{\textsf{sim}})=1\)) and then prefers minimal-change targets that are easier to realize in Stage~II (e.g., less perturbation).

Starting from an original compressed prompt \(x^{(0)}=[w_1,\dots,w_n]\) that exhibits \emph{normal} behavior after compression, we first use a small LLM to find critical words and greedily delete these words and re-evaluate the simulator with the judge after each deletion.     
Let \(x^{(0)}_{\setminus i}\) denote \(x^{(0)}\) with word \(w_i\) removed.
We iteratively apply deletions until the behavior flips to misbehavior, meaning \(J\!(y_{\textsf{sim}})=1\), and stop at the first such flip to keep the number of deletions small.
We denote by \(\mathcal{W}\) the set of deleted tokens up to the flip, and treat them as \emph{decision-critical tokens} that
decide the backend behavior (their removal induces misbehavior). The resulting compressed prompt
\(\tilde{x}_{\textsf{tgt}} \triangleq x^{(0)}_{\setminus \mathcal{W}}\) is used as the Stage~II target, where
we optimize a small perturbation so that the compressor drops \(\mathcal{W}\) and outputs \(\tilde{x}_{\textsf{tgt}}\).

\mypara{Stage II: Adversarial perturbation search}
Given the target prompt \(\tilde{x}_{\textsf{tgt}}\) in Stage I, we search for an attacked input \(x_{\textsf{atk}}\in\mathcal{N}(x)\) whose
surrogate-compressed prompt matches the target. For each \(\rho \in \mathcal{R}_{\textsf{cand}}\), we define a
compressed-space matching objective using the surrogate compressor's logits.

Let \(\tilde{x}_{\textsf{tgt}} = (\tilde{t}_1,\ldots,\tilde{t}_m)\). When running \(\hat{C}_{\rho}\) on a candidate
input \(x'\), the surrogate produces token logits \(z_j(x';\rho)\in\mathbb{R}^{|\mathcal{V}|}\) for the \(j\)-th
compressed token. 
We define
\[
\mathcal{L}_{\textsf{match}}(x';\rho)
\triangleq
-\sum_{j=1}^{m}\log \mathrm{softmax}\!\big(z_j(x';\rho)\big)[\tilde{t}_j],
\]
and aim to minimize \(\mathcal{L}_{\textsf{match}}(x';\rho)\) over \(x'\in\mathcal{N}(x)\).
We approximately solve this discrete optimization via greedy token replacement like GCG~\cite{zou2023universaltransferableadversarialattacks}.
Concretely, starting from \(x^{(0)}=x\), at each
iteration \(k\) we compute gradients of \(\mathcal{L}_{\textsf{match}}(x^{(k)};\rho)\) with respect to the embeddings
of editable tokens. For replacing token \(t_i\) by \(v\), we score the update with a first-order approximation
\(
\Delta_{i\rightarrow v}
\approx
\big(e_v - e_{t_i}\big)^{\top}\nabla_{e_{t_i}}\,\mathcal{L}_{\textsf{match}}(x^{(k)};\rho),
\)
and greedily apply the replacement \((i^\star,v^\star)\) that most decreases the objective while respecting
\(\mathcal{N}(x)\). 
We stop when \(\hat{C}_{\rho}(x^{(k)})\) matches \(\tilde{x}_{\textsf{tgt}}\) or exhaust the search budget, and keep the best candidate as \(x_{\textsf{atk}}(\rho)\).

In deployments where the compressor operates on an unobserved trusted input \(p\) (e.g., a system prompt) concatenated with the attacker-controlled untrusted input, the exact compressed-space behavior is unknown to the attacker.
We optimize a \emph{transfer} suffix \(\delta\) over an ensemble of surrogate prefix prompts \(\mathcal{P}_{\textsf{cand}}\), and then transfer this suffix to the hidden victim system prompt during final black-box validation.
We apply the same greedy replacement rule using gradients of \(\mathcal{L}_{\textsf{univ}}\):
\[
\mathcal{L}_{\textsf{univ}}(\delta)
\triangleq
\frac{1}{|\mathcal{P}_{\textsf{cand}}|}
\sum_{p\in\mathcal{P}_{\textsf{cand}}}
\mathcal{L}_{\textsf{match}}(p\Vert (x\oplus\delta);\rho).
\]

\mypara{End-to-End Validation}
All final results are validated end-to-end on the true black-box compressor \(C_R\) and backend LLM \(M\).
Specifically, we submit each candidate \(x'\) to the victim pipeline, obtain only the externally visible output transcript \(y'\), and evaluate success using the same binary judge \(J(y') \in \{0,1\}\).
Because the judge is binary, we select the \emph{least-modified} successful candidate:
\(
x_{\textsf{atk}} \in
\arg\min_{x' \in \mathcal{X}_{\textsf{cand}}}
\mathcal{D}(x',x),
\)
where \(J\!\big(y'\big)=1\), and \(\mathcal{X}_{\textsf{cand}}\) aggregates candidates across targets and
\(\rho \in \mathcal{R}_{\textsf{cand}}\), and \(\mathcal{D}\) is a modification measure, such as edit distance.
If no candidate achieves \(J(\cdot)=1\), we report failure and use the closest-to-target candidates for analysis.

\section{Experimental Setup}

\label{sec:exp_setup}
\label{sec:rq}
In this work, we aim to answer the following research questions:
\begin{itemize}[leftmargin=1.2em, topsep=2pt, itemsep=0pt, parsep=0pt, partopsep=0pt]
    \item \textbf{RQ1 (Effectiveness):} How effective is \ToolName{} compared to the non-compression-based attack baselines?

\item \textbf{RQ2 (Generalization):} How well does \ToolName{} generalize across different compression budgets \(R\) and backend LLMs?

\item  \textbf{RQ3 (Ablation):} How does the choice of surrogate model affect \ToolName{}'s testing effectiveness, and how does \ToolName{} steer information retention after compression?

\item \textbf{RQ4 (Judge Readability):} Is the LLM judge used to assess attack success reliable?

\item \textbf{RQ5: (Case Studies):} Can \ToolName{} transfer to realistic agent pipelines?

\item \textbf{RQ6 (Mitigation):} To what extent do existing mitigation methods remain effective against \ToolName{}?
\end{itemize}

\subsection{Tasks and Datasets}
\label{sec: data setup}
We follow the attack goals defined in threat model (\Cref{subsec:tm_goal}) with three tasks: two \emph{Evidence Distortion}
tasks under external-content perturbations \(\mathcal{N}_{\textsf{ctx}}(x)\), including \emph{Agent Tool selection (ATS)} and \emph{Question Answering (Q\&A)}, and one \emph{Guardrail Attenuation} task under budgeted query perturbations \(\mathcal{N}_{\textsf{que}}^{B}(x)\) which is \emph{System Prompt Corruption (SPC)}.
For each task, we report the macro-average over task-specific ASR metrics on task-filtered evaluation sets.

\noindent \ding{192} \mypara{Agent Tool Selection (ATS)}
We construct a dataset with 249 entries spanning 172 product categories and 77 tool categories; each entry contains five candidate target descriptions.
For each data entry, we first collect the initial tool descriptions from the Apify store~\cite{apify-store}, and leverage few-shot augmentation to generate the other 4 descriptions that belong to the same tool category with GPT-4o-mini generations.
Given a candidate set (with five candidates in our dataset), the backend LLM
outputs a top-1 choice \(y \in \mathcal{T}\). 
Let \(y_k^{*}\) denote the gold tool for instance \(k\), \(y_k^{\mathrm{clean}}\) the tool selected from the clean input, and \(y_k^{\mathrm{adv}}\) the tool selected from the attacked input.
We define ASR as the fraction of instances that are correct under the clean input but become incorrect after attack:
\(
\mathrm{ASR}_{\textsf{ATS}}
=
\sum_{k=1}^{N}
J\!\left[
y_k^{\mathrm{clean}} = y_k^{*}
\land
y_k^{\mathrm{adv}} \neq y_k^{*}
\right]\  / \
\sum_{k=1}^{N}
J\!\left[
y_k^{\mathrm{clean}} = y_k^{*}
\right]
.
\)

\noindent \ding{193} \textbf{Question Answering (Q\&A)}
We evaluate on SQuAD~\cite{rajpurkar-etal-2016-squad}, where each instance includes a Wikipedia
passage, a question, and the ground-truth answer. The passage is treated as untrusted external context
\(x^{\textsf{ctx}}\).
We sample 1000 data entries and filter the instance which the backend LLM cannot answer correctly without an attack to get 913 data entries.
The attacker aims to induce an incorrect answer.
Let \(y_k^\star\) be the ground-truth answer. We treat an attack as successful if the post-attack answer is incorrect under normalized exact match:
\(
\mathrm{ASR}_{\text{QA}}=\frac{1}{N}\sum_{k=1}^{N}J\!\left[
y_k^{\text{adv}}\neq y_k^\star
\right].
\)

\noindent \ding{194} \textbf{System Prompt Corruption (SPC)}
We collect a system-prompt dataset from four GitHub repositories, which collect the leaked system prompts from real-world agents~\cite{asgeirtjsystempromptsleaks,jujumilk3leakedsystemprompts,x1xhlolsystempromptsandmodelsofaitools,0xebTheBigPromptLibrary}, and the RealGuardrails benchmark~\cite{mu2025closer}, where each entry consists of a system prompt, a guardrail list, and a guardrail-violation query.
We use keyword matching to extract the safety-critical guardrails (e.g., don't, under no circumstances) and prompt GPT-4o-mini to generate the corresponding guardrail-violation query.
We then apply two screening filters to ensure each instance is informative for evaluating system-prompt corruption.
Using the semantic judge \(J(y)\in\{0,1\}\) (1=violation), we discard
(i) \emph{trivially vulnerable} instances where the clean run already violates the system prompt,
i.e., \(J(y_k^{\text{orig}})=1\); and
(ii) \emph{infeasible} instances where the backend still refuses the query even after removing the guardrail instructions from the system prompt, i.e., \(J(y_k^{-\textsf{sys}})=0\).
We keep only instances satisfying \(J(y_k^{\text{orig}})=0\) and \(J(y_k^{-\textsf{sys}})=1\), so that ASR reflects corruption of the system prompt rather than pre-existing non-compliance or model-level refusal.
Finally, we get 1563 data entries.
Let \(y_k^{\text{adv}}\) denote the backend output transcript returned by the black-box victim under attack. We report \(ASR_{\textsf{SPC}} = \frac{1}{N}\sum_{k=1}^{N} J(y_k^{\text{adv}})\).

To ensure meeting the length constraints of compressors and a fair comparison, we filter the dataset to include only entries with \(\leq 2048\) tokens (\(\leq 512\) for LLMLingua2 due to its BERT-architecture limitation).
Also, we filter all the data entries that fail without an attack to ensure the ASR is gained by the attack instead of the backend LLM bias and the compression bias.

\subsection{Surrogate Construction}
\label{sec:surrogate_construction}
\mypara{Testing Compressors}
To broadly cover prompt-compression mechanisms, we evaluate three representative \emph{extractive} compressors:
LLMLingua~\cite{jiang2023llmlinguacompressingpromptsaccelerated},
LLMLingua2~\cite{pan2024llmlingua2datadistillationefficient}, and
Selective Context (SC)~\cite{li2023compressing},
and an \emph{abstractive}, instruction-based summarization compressor instantiated with
three widely-used small LLM backbones: Qwen3-4B~\cite{yang2025qwen3}, Llama-3.2-3B~\cite{llama3modelcard}, and
Gemma-3-4B~\cite{team2024gemma}.
The semantic judge LLM $J$ used in this work is Qwen3-235B-A22B-Instruct-2507-FP8.
We use Llama-3.1-8B-instruct~\cite{llama3modelcard} as the default surrogate backend LLM.
We define the compression budget \(R\) as the fraction of tokens retained after compression. 
Thus, a smaller \(R\) corresponds to a tighter budget.
We set the candidate budget \(\mathcal{R}_{\textsf{cand}}\) defined in \Cref{subsec:method_two_stage} during optimization as \{0.5, 0.6, 0.7\}, which can largely preserve the semantic content of the original prompt as the previous paper reports~\cite{zakazov2025cmprsr}, and set the maximum optimization step to 500.
To keep perturbations small, we cap the perturbation budget at 32 tokens, which is \(1.5625\%\) of an overall 2048 token length.

\mypara{Surrogate Compressor Construction}
Our threat model assumes black-box access to the practical compressor \(C_R\)
To instantiate this setting, we construct a surrogate \(\hat{C}_{R}\) whose purpose is to approximate the \emph{compression decision primitive} of \(C_R\).
Compression is a selection policy, \ToolName{} requires a proxy of this policy to transfer attacks against the true black-box compressor.
We instantiate \(\hat{C}_{R}\) based on whether the target compressor is \emph{extractive} or \emph{abstractive}:
\begin{itemize}[label=\(\circ\), leftmargin=1.2em, topsep=1pt, itemsep=0pt, parsep=0pt, partopsep=0pt]
\item \textbf{Extractive compressors.}
Extractive compressors produce a subsequence of the original input by selecting tokens/spans under a budget \(R\).
We cover two common design patterns and build a surrogate accordingly:

\emph{(i) Scoring-based extractive compressors (Selective Context, LLMLingua).}
These compressors rank tokens/spans using predictability signals (e.g., perplexity) and retain the
highest-ranked content under budget \(R\).
Accordingly, we approximate the ranking signal with an accessible language model \(M_{\textsf{ppl}}\), compute token- or
span-level surprisal scores, and apply the same budgeted selection rule.
Crucially, the attack optimization depends on the \emph{relative ordering} of retained vs.\ discarded content, which is
typically stable across reasonable choices of \(M_{\textsf{ppl}}\), enabling transfer to the true \(C_R\).
We instantiate the surrogate compressor with Llama-2-7B in this work.

\emph{(ii) Learned extractive compressors (LLMLingua2).}
This kind of compressor is driven by a learned keep/drop classifier; we train a classifier-style surrogate.
Concretely, we construct weak supervision on MeetingBank~\cite{hu2023meetingbank} by generating length-controlled summaries with GPT-4o-mini~\cite{hurst2024gpt} and aligning them back to source tokens to obtain pseudo retention labels.
We then fine-tune a transformer token-classification model as \(\hat{C}_R\) to predict \(p(z_i=1\mid x,R)\) and enforce budget \(R\) via top-\(k\) selection. 
This matches the compressor's underlying decision primitive and provides guidance for gradients.
We instantiate the token-classification backbone using an encoder-style Transformer model, which is xlm-roberta-large.

\item \textbf{Abstractive compressors.}
For abstractive compressors, token-level retention is not well-defined due to paraphrasing. We therefore instantiate
\(\hat{C}_{R}\) as an accessible abstractive summarizer via a smaller LLM configured with the same length control, and optimize attacks against this surrogate in a transfer setting.
We use Qwen3-4B as the default surrogate abstractive compressor.
\end{itemize}
\section{Evaluation}

\subsection{RQ1: Effectiveness of \ToolName{}}
\label{sec:rq1}

In this RQ, we evaluate \ToolName{} under prompt compression on three tasks defined in \Cref{sec: data setup}.
To clarify whether \ToolName{} is genuinely \emph{compression-aware}, rather than simply acting as a generic malicious perturbation, we isolate the risk introduced specifically by the compressor and additionally report the ASR of \ToolName{} on the corresponding uncompressed pipeline (No Comp. + \ToolName{}), together with No-Attack ASR for the clean compressed pipeline.

To the best of our knowledge, no prior attack explicitly targets prompt compression.
We therefore compare against representative non-compression-based attacks to induce the misbehavior of backend LLM with the same goal as our attack, and apply these attacks to the same pre-compression input, and then passed through the same compression pipeline.
\ding{172} Naive Attack: random character-level perturbations of selected words (budget-matched).
\ding{173} Direct Injection~\cite{liu2024formalizing}: append an explicit malicious instruction (budget-matched).
\ding{174} Escape Characters~\cite{willison2022prompt}: insert separators (e.g., \texttt{\textbackslash n}, \texttt{\textbackslash t}) before the malicious instruction.
\ding{175} Context Ignore~\cite{branch2022evaluatingsusceptibilitypretrainedlanguage,perez2022ignorepreviouspromptattack}: insert override phrases (e.g., ``ignore previous instructions/context'').
\ding{176} JudgeDeceiver~\cite{shi2025optimizationbasedpromptinjectionattack}: 
This is a gradient-based adversarial prompts generation method.
We consider a \emph{transfer} setting of this attack because our attack setting is black-box.

\newcommand{\best}[1]{\textbf{#1}}
\newcommand{\bestbase}[1]{\underline{#1}}

\begin{table}[!th]
\centering
\scriptsize
\setlength{\tabcolsep}{1.6pt}
\renewcommand{\arraystretch}{0.75}

\begin{threeparttable}
\caption{\small
Attack success rate (ASR, \(\uparrow\)) of \ToolName{} and non-compression-based attack baselines under six prompt compressors on three tasks: ATS, (Q\&A), and SPC. 
In the \textbf{Task Avg.} block, the number in parentheses after each task name denotes the corresponding no-compression ASR of \ToolName{} for that task. 
}
\label{tab:asr_by_compressor_cols_baselines}

\begin{tabular}{llccccccc}
\toprule
\multirow{2}{*}{\textbf{Compressor}} & \multirow{2}{*}{\textbf{Task}} &
\multirow{2}{*}{\textbf{No Attack}} & \multirow{2}{*}{\textbf{\ToolName{}}} &
\multicolumn{5}{c}{\textbf{Baselines}} \\
\cmidrule(lr){5-9}
& & & &
\textbf{Naive} &
\shortstack{\textbf{Direct}\\\textbf{Injection}} &
\shortstack{\textbf{Escape}\\\textbf{Chars}} &
\shortstack{\textbf{Context}\\\textbf{Ignore}} &
\shortstack{\textbf{Judge}\\\textbf{Deceiver}} \\
\midrule

\multicolumn{9}{@{}l}{\textit{Extractive Compressors}}\\

\multirow{3}{*}{SC}
& ATS  & 0.00 & \bestcell{0.98} & 0.36 & 0.38 & \secondcell{0.41} & 0.38 & 0.26 \\
& Q\&A & 0.00 & \bestcell{0.80} & 0.19 & 0.22 & \secondcell{0.24} & 0.21 & 0.17 \\
& SPC  & 0.00 & \bestcell{0.62} & 0.00 & 0.04 & 0.06 & \secondcell{0.34} & 0.12 \\

\multirow{3}{*}{LLMLingua}
& ATS  & 0.00 & \bestcell{0.96} & 0.31 & 0.34 & \secondcell{0.36} & 0.34 & 0.22 \\
& Q\&A & 0.00 & \bestcell{0.62} & 0.15 & 0.16 & \secondcell{0.19} & 0.18 & 0.15 \\
& SPC  & 0.00 & \bestcell{0.52} & 0.00 & 0.01 & 0.02 & \secondcell{0.30} & 0.18 \\

\multirow{3}{*}{LLMLingua2}
& ATS  & 0.00 & \bestcell{0.98} & 0.24 & \secondcell{0.30} & 0.23 & 0.29 & 0.20 \\
& Q\&A & 0.00 & \bestcell{0.72} & 0.04 & 0.08 & 0.18 & \secondcell{0.24} & 0.09 \\
& SPC  & 0.00 & \bestcell{0.63} & 0.00 & 0.00 & 0.10 & \secondcell{0.34} & 0.20 \\

\midrule
\multicolumn{9}{@{}l}{\textit{Abstractive Compressors}}\\

\multirow{3}{*}{Llama-3.2-3B}
& ATS  & 0.04 & \bestcell{0.86} & 0.02 & 0.07 & 0.10 & 0.14 & \secondcell{0.18} \\
& Q\&A & 0.00 & \bestcell{0.69} & 0.00 & 0.00 & 0.00 & 0.05 & \secondcell{0.32} \\
& SPC  & 0.04 & \bestcell{0.54} & 0.00 & 0.00 & 0.07 & 0.09 & \secondcell{0.18} \\

\multirow{3}{*}{Qwen3-4B}
& ATS  & 0.01 & \bestcell{0.82} & 0.02 & 0.09 & 0.05 & 0.09 & \secondcell{0.16} \\
& Q\&A & 0.00 & \bestcell{0.59} & 0.00 & 0.00 & 0.00 & 0.02 & \secondcell{0.32} \\
& SPC  & 0.02 & \bestcell{0.49} & 0.00 & 0.00 & 0.06 & 0.14 & \secondcell{0.22} \\

\multirow{3}{*}{Gemma-3-4B}
& ATS  & 0.02 & \bestcell{0.89} & 0.04 & 0.06 & 0.08 & \secondcell{0.18} & 0.16 \\
& Q\&A & 0.00 & \bestcell{0.71} & 0.00 & 0.00 & 0.01 & 0.04 & \secondcell{0.36} \\
& SPC  & 0.04 & \bestcell{0.44} & 0.00 & 0.01 & 0.04 & 0.12 & \secondcell{0.21} \\

\midrule

\rowcolor{avggray}
\multirow{3}{*}{\textbf{Task Avg.}}
& ATS \textbf{(0.01)} & \avgcell{0.01} & \avgcell{\bestcell{0.92}} & \avgcell{0.17} & \avgcell{0.21} & \avgcell{0.21} & \avgcell{\secondcell{0.24}} & \avgcell{0.20} \\
\rowcolor{avggray}
& Q\&A \textbf{(0.00)} & \avgcell{0.00} & \avgcell{\bestcell{0.69}} & \avgcell{0.06} & \avgcell{0.08} & \avgcell{0.10} & \avgcell{0.12} & \avgcell{\secondcell{0.24}} \\
\rowcolor{avggray}
& SPC  \textbf{(0.00)} & \avgcell{0.02} & \avgcell{\bestcell{0.54}} & \avgcell{0.00} & \avgcell{0.01} & \avgcell{0.06} & \avgcell{\secondcell{0.22}} & \avgcell{0.19} \\
\midrule

\rowcolor{avggray}
\multicolumn{2}{l}{\textbf{Avg (Extractive)}} & \avgcell{0.00} & \avgcell{\bestcell{0.76}} & \avgcell{0.14} & \avgcell{0.17} & \avgcell{0.20} & \avgcell{\secondcell{0.29}} & \avgcell{0.18} \\
\rowcolor{avggray}
\multicolumn{2}{l}{\textbf{Avg (Abstractive)}} & \avgcell{0.02} & \avgcell{\bestcell{0.67}} & \avgcell{0.01} & \avgcell{0.03} & \avgcell{0.05} & \avgcell{0.10} & \avgcell{\secondcell{0.23}} \\
\rowcolor{avggray}
\multicolumn{2}{l}{\textbf{Overall Avg.}} & \avgcell{0.01} & \avgcell{\bestcell{0.71}} & \avgcell{0.08} & \avgcell{0.10} & \avgcell{0.12} & \avgcell{0.19} & \avgcell{\secondcell{0.21}} \\
\bottomrule
\end{tabular}

\end{threeparttable}
\end{table}

\mypara{Overall Performance}
\Cref{tab:asr_by_compressor_cols_baselines} summarizes the ASR of all methods across six compressors and three tasks. \ToolName{} achieves the highest ASR in every setting (\(18/18\)). On average, \ToolName{} reaches \(0.71\) ASR, compared with \(0.21\) for the strongest baseline, a gain of \(+0.50\) absolute points (about \(3.4\times\)).
The No-Attack condition yields an overall ASR of only \(0.01\), suggesting that the compressed pipeline itself rarely triggers the target misbehaviors in the absence of an adversarial input. 
And the no-compression control for \ToolName{} is also near zero overall (\(0.01\)), indicating that \ToolName{} is ineffective when the compressor is removed, and explains that \ToolName{} derives its effectiveness from \Vun{} vulnerabilities introduced by the prompt-compression stage.
The reason why non-compression-based baselines underperform is that they require their malicious payloads to survive compression and remain interpretable by the backend LLM, whereas COMA uses perturbations to steer retention decisions and therefore does not require the payload itself to survive compression.

\mypara{Task-wise Susceptibility}
\ToolName{} is consistently effective across all three tasks, but the degree of vulnerability depends on whether the attack directly perturbs compressor-visible external content or must rely on suffix transfer.
ATS and Q\&A are more vulnerable because the adversary modifies external content that is directly compressed, making it easier to alter what survives to the backend LLM.
SPC task is harder because the attack depends on the transferability of the suffix perturbation to the real system prompt, which reduces reliability.
\begin{Result}
\textbf{\ToolName{} consistently achieves the highest ASR across all \(18/18\) settings.} The near-zero ASR of both \textsc{No-Attack} and the no-compression control shows that its success comes from exploiting vulnerabilities of prompt compression.
\end{Result}

\subsection{RQ2: Generalization of \ToolName{}}
In this RQ, we study the generalization of \ToolName{} along two axes. 
First, we sweep the compression budget to vary the strength of the information bottleneck.
Second, we fix the compressor and evaluate multiple backend LLMs from different families and scales to test generalization across backend LLMs.

\begin{table}[!ht]
\centering
\scriptsize
\setlength{\tabcolsep}{0.1pt}
\renewcommand{\arraystretch}{0.95}
\caption{\small Attack success rate (ASR, $\uparrow$) of \ToolName{} across the compression budget \(R \in \{0.2,0.4,0.6,0.8\}\).
Each cell is reported as \(a_{\Delta d}\), where \(a\) is the ASR under attack and \(\Delta d=a-\text{ASR}_{\text{NoAttack}}\) is the attack-induced increase beyond benign compression at the same budget.}
\label{tab:rq_budget_sweep_qa}
\resizebox{\linewidth}{!}{
\begin{tabular}{lcccccccccc}
\toprule
\multirow{2}{*}{Compressor}
& \multicolumn{2}{c}{\(R=0.2\)}
& \multicolumn{2}{c}{\(R=0.4\)}
& \multicolumn{2}{c}{\(R=0.6\)}
& \multicolumn{2}{c}{\(R=0.8\)}
& \multicolumn{2}{c}{\avgcell{\textbf{Avg.}}} \\
\cmidrule(lr){2-3}\cmidrule(lr){4-5}\cmidrule(lr){6-7}\cmidrule(lr){8-9}\cmidrule(lr){10-11}
& ASR & No-attack & ASR & No-attack & ASR & No-attack & ASR & No-attack & \avgcell{ASR} & \avgcell{No-attack} \\
\midrule

SC
& \bestcell{\(\mathbf{0.86}_{\Delta \mathbf{0.03}}\)} & \bestcell{0.83}
& \(0.64_{\Delta0.28}\) & \secondcell{0.36}
& \bestcell{\(\mathbf{0.80}_{\Delta \mathbf{0.80}}\)} & 0.00
& \(0.66_{\Delta0.66}\) & 0.00
& \bestcell{\(\mathbf{0.74}_{\Delta \mathbf{0.44}}\)} & \bestcell{0.30} \\

LLMLingua
& \secondcell{\(0.79_{\Delta0.15}\)} & \(0.64\)
& \(0.54_{\Delta0.28}\) & \(0.26\)
& \(0.62_{\Delta0.62}\) & 0.00
& \(0.62_{\Delta0.62}\) & 0.00
& \avgcell{\(0.64_{\Delta0.41}\)} & \avgcell{0.23} \\

LLMLingua2
& \secondcell{\(0.79_{\Delta0.31}\)} & \(0.48\)
& \bestcell{\(\mathbf{0.79}_{\Delta \mathbf{0.57}}\)} & \(0.22\)
& \secondcell{\(0.72_{\Delta0.72}\)} & 0.00
& \secondcell{\(0.67_{\Delta0.67}\)} & 0.00
& \bestcell{\(\mathbf{0.74}_{\Delta \mathbf{0.56}}\)} & \avgcell{0.18} \\

Llama-3.2-3B
& \(0.64_{\Delta0.05}\) & \(0.59\)
& \secondcell{\(0.74_{\Delta0.35}\)} & \bestcell{0.39}
& \(0.69_{\Delta0.67}\) & \secondcell{0.02}
& \bestcell{\(\mathbf{0.72}_{\Delta \mathbf{0.72}}\)} & 0.00
& \avgcell{\(0.70_{\Delta0.45}\)} & \avgcell{0.25} \\

Qwen3-4B
& \(0.56_{\Delta0.14}\) & \(0.42\)
& \(0.63_{\Delta0.49}\) & \(0.14\)
& \(0.59_{\Delta0.59}\) & 0.00
& \(0.55_{\Delta0.55}\) & 0.00
& \avgcell{\(0.58_{\Delta0.44}\)} & \avgcell{0.14} \\

Gemma-3-4B
& \(0.78_{\Delta0.12}\) & \secondcell{0.66}
& \(0.68_{\Delta0.34}\) & \(0.34\)
& \(0.71_{\Delta0.67}\) & \bestcell{0.04}
& \(0.65_{\Delta0.64}\) & \bestcell{0.01}
& \secondcell{\(0.71_{\Delta0.45}\)} & \secondcell{0.26} \\

\midrule
\rowcolor{avggray}
\textbf{Avg.}
& \(\mathbf{0.74}_{\Delta \mathbf{0.14}}\) & \bestcell{\textbf{0.60}}
& \(\mathbf{0.67}_{\Delta \mathbf{0.38}}\) & \secondcell{\textbf{0.29}}
& \bestcell{\(\mathbf{0.69}_{\Delta \mathbf{0.68}}\)} & \textbf{0.01}
& \secondcell{\(\mathbf{0.65}_{\Delta \mathbf{0.65}}\)} & \textbf{0.00}
& \textbf{\(\mathbf{0.69}_{\Delta \mathbf{0.46}}\)} & \textbf{0.23} \\
\bottomrule
\end{tabular}
}
\end{table}
\mypara{Generalization across Budgets}
\Cref{tab:rq_budget_sweep_qa} reports the ASR under \ToolName{} and no-attack setting at a victim budget \(R \in\{0.2, 0.4, 0.6, 0.8\}\) in the Q\&A task.
\label{sec:rq2_budget}
\emph{\ToolName{} generalizes well across different compression budgets, achieving a high average ASR of 0.69.}
Averaged over compressors in \Cref{tab:rq_budget_sweep_qa}, the no-attack ASR is \(0.60\) at \(R=0.2\) and \(0.29\) at \(R=0.4\), indicating that tight budgets can collapse many task-relevant distinctions even in benign settings. This trend is consistent with prior work~\cite{zakazov2025cmprsr} and suggests that overly small compression budgets should be avoided.
By contrast, the no-attack ASR becomes nearly zero at \(R\ge 0.6\) (\(0.01\) at \(R=0.6\) and \(0.00\) at \(R=0.8\)), suggesting that these budgets largely preserve the information needed for correct Q\&A. However, adversarial ASR remains consistently high: even at \(R\in\{0.6,0.8\}\), where the benign pipeline rarely fails, \ToolName{} still induces failures in a large fraction of instances, with ASRs of \(0.69\) and \(0.65\), respectively.

\emph{Attack amplification peaks at deployment-relevant budgets.}
To isolate attacker-amplified distortion beyond compression alone, we report the attack-benign gap (ASR under attack minus no-attack ASR) at the same budget to present the amplification effectiveness of \ToolName{}.
From the averaged row of \Cref{tab:rq_budget_sweep_qa}, this gap increases from 0.14 at \(R=0.2\) to 0.38 at \(R=0.4\), peaks at 0.68 at \(R=0.6\), and remains high at 0.65 for \(R=0.8\).
Amplification is therefore largest for \(R\ge 0.6\): the benign system is highly accurate, yet an attacker can still reliably induce failures.
However, at \(R=0.2\), benign compression is already highly destructive (no-attack ASR = 0.60), leaving limited room for adversarial amplification.

\begin{table}[!ht]
\setlength{\tabcolsep}{2.2pt}
\scriptsize
\renewcommand{\arraystretch}{0.75}
\centering
\caption{\small Attack success rate (ASR, \(\uparrow\)) of \ToolName{} across backend LLM families and model scales on ATS, Q\&A, and SPC tasks across six backend LLMs.}
\label{tab:preference_across_models_token_only}
\begin{tabular}{llccccccc}
\toprule
\multirow{3}{*}{\textbf{Compressor}} &
\multirow{3}{*}{\textbf{Task}} &
\multicolumn{6}{c}{\textbf{Backend LLM (ASR, \(\uparrow\))}} &
\multirow{3}{*}{\textbf{Avg.}} \\
\cmidrule(lr){3-8}
& & \shortstack{\textbf{Llama-2}\\\textbf{-7B}} & \shortstack{\textbf{Llama-3}\\ \textbf{-8B}} & \shortstack{\textbf{Llama-3}\\ \textbf{-70B}} & \shortstack{\textbf{Mistral-2}\\ \textbf{-7B}} & \shortstack{\textbf{Qwen3}\\ \textbf{-14B}} & \shortstack{\textbf{Qwen3}\\ \textbf{-32B}} & \\
\midrule

\multirow{3}{*}{SC}
& ATS   & 0.95 & 0.98 & 0.91 & 0.95 & 0.95 & 0.94 & \avgcell{\textbf{0.95}} \\
& Q\&A  & 0.84 & 0.80 & 0.78 & 0.83 & 0.49 & 0.41 & \avgcell{\textbf{0.69}} \\
& SPC   & 0.78 & 0.62 & 0.55 & 0.60 & 0.56 & 0.50 & \avgcell{\textbf{0.60}} \\
\midrule

\multirow{3}{*}{LLMLingua}
& ATS   & 0.96 & 0.96 & 0.95 & 0.96 & 0.94 & 0.94 & \avgcell{\textbf{0.95}} \\
& Q\&A  & 0.78 & 0.62 & 0.60 & 0.75 & 0.45 & 0.41 & \avgcell{\textbf{0.60}} \\
& SPC   & 0.62 & 0.52 & 0.49 & 0.57 & 0.48 & 0.47 & \avgcell{\textbf{0.53}} \\
\midrule

\multirow{3}{*}{LLMLingua2}
& ATS   & 0.98 & 0.98 & 0.97 & 0.98 & 0.95 & 0.95 & \avgcell{\textbf{0.97}} \\
& Q\&A  & 0.77 & 0.72 & 0.67 & 0.78 & 0.54 & 0.45 & \avgcell{\textbf{0.66}} \\
& SPC   & 0.74 & 0.63 & 0.60 & 0.71 & 0.59 & 0.53 & \avgcell{\textbf{0.63}} \\
\midrule

\multirow{3}{*}{Llama-3.2-3B}
& ATS   & 0.86 & 0.86 & 0.86 & 0.85 & 0.86 & 0.85 & \avgcell{\textbf{0.86}} \\
& Q\&A  & 0.72 & 0.69 & 0.66 & 0.73 & 0.51 & 0.46 & \avgcell{\textbf{0.63}} \\
& SPC   & 0.59 & 0.54 & 0.51 & 0.60 & 0.49 & 0.47 & \avgcell{\textbf{0.53}} \\
\midrule

\multirow{3}{*}{Qwen3-4B}
& ATS   & 0.80 & 0.82 & 0.82 & 0.82 & 0.80 & 0.82 & \avgcell{\textbf{0.81}} \\
& Q\&A  & 0.63 & 0.59 & 0.58 & 0.65 & 0.49 & 0.42 & \avgcell{\textbf{0.56}} \\
& SPC   & 0.60 & 0.49 & 0.48 & 0.58 & 0.46 & 0.39 & \avgcell{\textbf{0.50}} \\
\midrule

\multirow{3}{*}{Gemma-3-4B}
& ATS   & 0.89 & 0.89 & 0.88 & 0.89 & 0.89 & 0.89 & \avgcell{\textbf{0.89}} \\
& Q\&A  & 0.78 & 0.71 & 0.69 & 0.76 & 0.59 & 0.55 & \avgcell{\textbf{0.68}} \\
& SPC   & 0.52 & 0.44 & 0.43 & 0.50 & 0.40 & 0.39 & \avgcell{\textbf{0.45}} \\
\midrule

\rowcolor{avggray}
\multicolumn{2}{c}{\textbf{Overall Avg.}} &
\bestcell{0.77} & 0.71 & 0.69 & \secondcell{0.75} &
0.64 & 0.60 & \avgcell{\textbf{0.69}} \\
\bottomrule
\end{tabular}
\end{table}

\mypara{Generalization across Backend LLMs}
\label{sec:rq2_backend}
Table~\ref{tab:preference_across_models_token_only} evaluates generalization across backend LLM families (Llama/Mistral/Qwen) and scales (7B-70B, 14B-32B) with fixed compressors.
Overall, \ToolName{} exhibits strong generalization: the average ASR ranges from \(0.60\) on Qwen3-32B to \(0.77\) on Llama-2-7B, with an overall mean of \(0.69\).
This consistent effectiveness across diverse backends suggests that \ToolName{} primarily exploits a structural vulnerability at the compression boundary.
Backend variation may affect how the model reasons over the compressed prompt, but it cannot recover task-critical information that has already been removed or distorted by compression.

\emph{Task-wise patterns further clarify the mechanism.}
ATS is near-saturated and highly stable across backends: averaging over compressors, its ASR ranges only from \(0.90\) to \(0.92\) across the six backend models. By contrast, Q\&A and SPC show substantially larger backend dependence. Their average ASR across compressors ranges from \(0.45\) to \(0.75\) for Q\&A and from \(0.46\) to \(0.64\) for SPC. 
This pattern arises because ATS success is driven almost entirely by compression-stage distortion, when the key sentiment evidence is dropped, any backend will flip its prediction. 
Whereas Q\&A and SPC depend heavily on the backend's own capabilities. 
A stronger backend LLM can fall back on its parametric knowledge to answer correctly, despite a corrupted context in Q\&A task, or rely on its stronger safety alignment to
uphold guardrails even when guardrails are stripped from the compressed prompt in SPC task.

\begin{Result}\label{result:rq2-generalization}
\textbf{\ToolName{} generalizes across both compression budgets and backend LLMs.}
In the budget sweep, attack success remains high even when the no-attack ASR is near zero.
Across backend families and scales, \ToolName{} generalizes consistently with an overall mean ASR of \(0.69\).
\end{Result}

\subsection{RQ3: Surrogates and Retention Steering}
\label{sec:rq3-surrogate}

In this RQ, we conduct a diagnostic ablation of \ToolName{} from two complementary perspectives.
First, to reveal how sensitive \ToolName{} is to the choice of surrogate model, we study \emph{surrogate transferability} by varying the attacker-side surrogate used to approximate a black-box compressor and measuring the resulting change in end-to-end attack success.
Second, we examine the attack mechanism itself by analyzing how \ToolName{} steers the compressor's information retention by evaluating whether it causes critical tokens to be discarded after compression.

\begin{table}[!ht]
\centering
\scriptsize
\setlength{\tabcolsep}{7pt}
\renewcommand{\arraystretch}{0.75}
\caption{\small Surrogate transferability under target-surrogate mismatch. 
We report attack success rate (ASR, \(\uparrow\)) on ATS, Q\&A, and SPC; Within each target-compressor block, the highest Avg. is boldfaced, and the remaining Avg. entries show the gap to the block maximum, \(\Delta=\mathrm{Avg.}-\mathrm{Avg.}_{\max}\).}
\label{tab:surrogate_mismatch_pre}
\begin{tabular}{llcccc}
\toprule
\textbf{Target Compressor} & \textbf{Surrogate Model}
& \textbf{ATS}
& \textbf{Q\&A}
& \textbf{SPC}
& \textbf{Avg.} \\
\midrule

\multirow{3}{*}{SC}
& GPT-2      & 0.80 & 0.70 & 0.51 &  \avgcell{\(0.67_{\Delta-0.13}\)} \\
& Llama-2-7B & 0.98 & 0.81 & 0.62 & \bestcell{\textbf{0.80}} \\
& Phi-2      & 0.94 & 0.75 & 0.58 & \avgcell{\(0.76_{\Delta-0.05}\)} \\
\midrule

\multirow{3}{*}{LLMLingua}
& GPT-2      & 0.78 & 0.55 & 0.39 & \avgcell{\(0.57_{\Delta-0.16}\)} \\
& Llama-2-7B & 0.96 & 0.69 & 0.54 & \bestcell{\textbf{0.73}} \\
& Phi-2      & 0.89 & 0.65 & 0.54 &  \avgcell{\(0.69_{\Delta-0.04}\)} \\
\midrule

\multirow{3}{*}{LLMLingua2}
& bert-base-uncased & 0.84 & 0.55 & 0.49 & \avgcell{\(0.63_{\Delta-0.15}\)} \\
& deberta-v3-large  & 0.93 & 0.59 & 0.52 & \avgcell{\(0.68_{\Delta-0.10}\)} \\
& xlm-roberta-large & 0.98 & 0.72 & 0.63 &  \bestcell{\textbf{0.78}} \\
\midrule

\multirow{3}{*}{Llama-3.2-3B}
& Llama-3.2-3B & 0.86 & 0.71 & 0.58 & \bestcell{\textbf{0.72}} \\
& Qwen3-4B     & 0.85 & 0.70 & 0.56 & \avgcell{\(0.70_{\Delta-0.01}\)} \\
& Gemma-3-4B   & 0.81 & 0.61 & 0.46 & \avgcell{\(0.63_{\Delta-0.09}\)} \\
\midrule

\multirow{3}{*}{Qwen3-4B}
& Llama-3.2-3B & 0.80 & 0.52 & 0.45 & \avgcell{\(0.59_{\Delta-0.05}\)} \\
& Qwen3-4B     & 0.82 & 0.58 & 0.51 & \bestcell{\textbf{0.64}} \\
& Gemma-3-4B   & 0.80 & 0.50 & 0.44 & \avgcell{\(0.58_{\Delta-0.06}\)} \\
\midrule

\multirow{3}{*}{Gemma-3-4B}
& Llama-3.2-3B & 0.85 & 0.69 & 0.45 & \avgcell{\(0.66_{\Delta-0.04}\)} \\
& Qwen3-4B     & 0.88 & 0.72 & 0.44 & \avgcell{\(0.68_{\Delta-0.02}\)} \\
& Gemma-3-4B   & 0.89 & 0.74 & 0.48 & \bestcell{\textbf{0.70}} \\
\bottomrule
\end{tabular}
\end{table}

\mypara{Surrogate Transferability Study}
We study whether \ToolName{} requires an exact replica of the deployed compressor. For Selective Context and LLMLingua, we instantiate the surrogate with GPT-2, Llama-2-7B, and Phi-2. 
For LLMLingua2, we vary the token-classification backbone among bert-base-uncased, deberta-v3-large, and xlm-roberta-large.
For abstractive compressors, we perform cross-surrogate transfer by treating each compressor as a surrogate for the others.
Due to resource limitations, we report results on 100 randomly sampled instances per dataset.

\emph{\Cref{tab:surrogate_mismatch_pre} shows that surrogate mismatch lowers ASR but does not eliminate transfer.}
Across the 12 mismatched settings, the mean ASR remains \(0.65\), compared with \(0.73\) for the best surrogate within each target-compressor block. For extractive compressors, better-aligned surrogates consistently perform best: Llama-2-7B achieves the highest average ASR on Selective Context (\(0.80\)) and LLMLingua (\(0.73\)), while XLM-R-large performs best on LLMLingua2 (\(0.78\)). Nevertheless, mismatched surrogates remain effective, e.g., Phi-2 still reaches \(0.76\) on Selective Context and \(0.69\) on LLMLingua.

\emph{These results suggest that exact replication of the victim compressor is unnecessary; a surrogate only needs to approximate what information survives compression.}
For abstractive compressors, the matched surrogate yields the best ASR in each block, but cross-surrogate transfer remains strong with only modest degradation (\(\Delta=-0.01\) to \(-0.09\) in Avg.). Across tasks, ATS is the most robust under surrogate mismatch, while Q\&A and especially SPC are more sensitive. 


\begin{table}[!ht]
\centering
\scriptsize
\setlength{\tabcolsep}{4pt}
\renewcommand{\arraystretch}{0.82}
\caption{\small  Critical Token Removal Rate (CTRR, \(\uparrow\)) and Attack Success Rate (ASR, \(\uparrow\)) on ATS, Q\&A, and SPC tasks across six compressors.}
\label{tab:compressor_ctrr_asr_camera_ready}
\begin{tabular}{lcccccccc}
\toprule
\multirow{2}{*}{\textbf{Target Compressor}}
& \multicolumn{2}{c}{\textbf{ATS}}
& \multicolumn{2}{c}{\textbf{Q\&A}}
& \multicolumn{2}{c}{\textbf{SPC}}
& \multicolumn{2}{c}{\avgcell{\textbf{Avg.}}} \\
\cmidrule(lr){2-3}\cmidrule(lr){4-5}\cmidrule(lr){6-7}\cmidrule(lr){8-9}
& \textbf{CTRR} & \textbf{ASR}
& \textbf{CTRR} & \textbf{ASR}
& \textbf{CTRR} & \textbf{ASR}
& \avgcell{\textbf{CTRR}} & \avgcell{\textbf{ASR}} \\
\midrule

\multicolumn{9}{l}{\textit{Extractive Compressors}} \\
SC & \secondcell{0.98} & \bestcell{0.98} & \bestcell{0.84} & \bestcell{0.80} & \secondcell{0.62} & \secondcell{0.62} & \avgcell{\bestcell{0.81}} & \avgcell{\bestcell{0.80}} \\
LLMLingua        & 0.97 & \secondcell{0.96} & 0.76 & 0.62 & 0.52 & 0.52 & \avgcell{0.75} & \avgcell{0.70} \\
LLMLingua2       & \bestcell{0.99} & \bestcell{0.98} & \secondcell{0.79} & \secondcell{0.72} & \bestcell{0.63} & \bestcell{0.63} & \avgcell{\secondcell{0.80}} & \avgcell{\secondcell{0.78}} \\
\rowcolor{gray!8}
\textbf{Extractive Avg.} & 0.98 & 0.97 & 0.80 & 0.71 & 0.59 & 0.59 & \textbf{0.79} & \textbf{0.76} \\
\midrule

\multicolumn{9}{l}{\textit{Abstractive Compressors}} \\
Llama-3.2-3B & 0.94 & 0.86 & 0.75 & 0.69 & 0.54 & 0.54 & \avgcell{0.74} & \avgcell{0.70} \\
Qwen3-4B     & 0.89 & 0.82 & 0.67 & 0.59 & 0.49 & 0.49 & \avgcell{0.68} & \avgcell{0.63} \\
Gemma-3-4B   & 0.96 & 0.89 & \bestcell{0.84} & 0.71 & 0.44 & 0.44 & \avgcell{0.75} & \avgcell{0.68} \\
\rowcolor{gray!8}
\textbf{Abstractive Avg.} & 0.93 & 0.86 & 0.75 & 0.66 & 0.49 & 0.49 & \textbf{0.72} & \textbf{0.67} \\
\midrule

\rowcolor{gray!15}
\textbf{Overall Avg.} & 0.96 & 0.92 & 0.78 & 0.69 & 0.54 & 0.54 & \textbf{0.76} & \textbf{0.72} \\
\bottomrule
\end{tabular}
\end{table}
\mypara{Critical Token Retention Study}
To quantify whether COMA successfully removes the decision-critical tokens identified in Stage~I (\Cref{subsec:method_two_stage}), we measure the Critical Token Removal Rate (CTRR). For each instance \(i\), let \(W_i\) denote the set of critical-token occurrences identified during target selection, and let \(\tilde{x}_i^{adv}\) be the attacked compressed prompt. 
We define
\(
\mathrm{CTRR}_i = 1 - \frac{1}{|W_i|} \sum_{w \in W_i} \mathbbm{1}[w \in \tilde{x}_i^{adv}].
\)
For extractive compressors, CTRR directly measures token removal.
For abstractive compressors, it serves only as an approximate proxy for the loss of the targeted content.
We report the macro-average CTRR over instances. A higher CTRR indicates that more targeted tokens are removed after compression.

\emph{Table~\ref{tab:compressor_ctrr_asr_camera_ready} shows that CTRR closely tracks ASR across compressors and tasks.}
The overall averages are 0.76/0.72, with the same pattern for extractive (0.79/0.76) and abstractive (0.72/0.67) compressors. This alignment suggests that COMA succeeds primarily by steering the compressor to discard the Stage-I critical tokens, rather than by merely appending generic malicious content.

\emph{Task-wise results further clarify the mechanism.}
ATS achieves the highest CTRR and ASR (0.96/0.92), consistent with tool selection depending on a small set of highly discriminative cues. 
Q\&A shows a larger CTRR--ASR gap (0.78/0.69). This is because answer-relevant evidence is more distributed and partially redundant, so the backend can sometimes recover from residual context. SPC shows an exact match (0.54/0.54), indicating a particularly direct risk: once safety-critical guardrail tokens are removed, the refusal condition will disappear.

\begin{Result}
\ToolName{} is robust to surrogate transferability.
Better-aligned surrogates yield the highest ASR, but mismatched surrogates still transfer effectively across compressors. 
And \ToolName{} works by reliably and controllably removing a small amount of behavior-critical information at the compression boundary
\end{Result}

\subsection{RQ4: Reliability of the LLM Judge Oracle}

\begin{table}[!ht]
\centering
\setlength{\tabcolsep}{4pt}
\renewcommand{\arraystretch}{0.75}
\scriptsize
\caption{\small Agreement between the LLM judge and human annotation on 300 stratified samples. AF denotes the Attack Fail class.}
\label{tab:llm_judge_faithfulness}
\begin{tabular}{lccccc}
\toprule
\textbf{Metric} & \textbf{Agreement (\%)} & \textbf{Cohen's \(\kappa\)} & \textbf{Precision\(_{\textsc{AF}}\) (\%)} & \textbf{Recall\(_{\textsc{AF}}\) (\%)} & \textbf{F1\(_{\textsc{AF}}\) (\%)} \\
\midrule
\textbf{Value}  & 98.7           & 0.95               & 93.3                        & 97.7                     & 95.5                 \\
\bottomrule
\end{tabular}%
\end{table}

In this RQ, we evaluate whether the LLM judge used to determine attack success is faithful to human evaluation.
We sample 300 instances, stratified by task and verdict, and manually annotate each instance as \textsc{Attack Success} or \textsc{Attack Fail}.
We then compare the LLM judge's predictions against these human labels, reporting the agreement rate, Cohen's $\kappa$, and precision/recall/F1 for the \textsc{Attack Fail} class.
As shown in \Cref{tab:llm_judge_faithfulness}, the LLM judge achieves 98.7\% agreement with human labels (Cohen's $\kappa$ = 0.95), with an F1 score of 95.5\% on the \textsc{Attack Fail} class, confirming that it is a reliable oracle for judging attack success.

\begin{Result}
The LLM judge is reliable for evaluating attack success, achieving 98.7\% agreement with human annotation.
\end{Result}
\subsection{RQ5: Case Studies}

\begin{figure}[!ht]
    \centering
    \includegraphics[width=\linewidth]{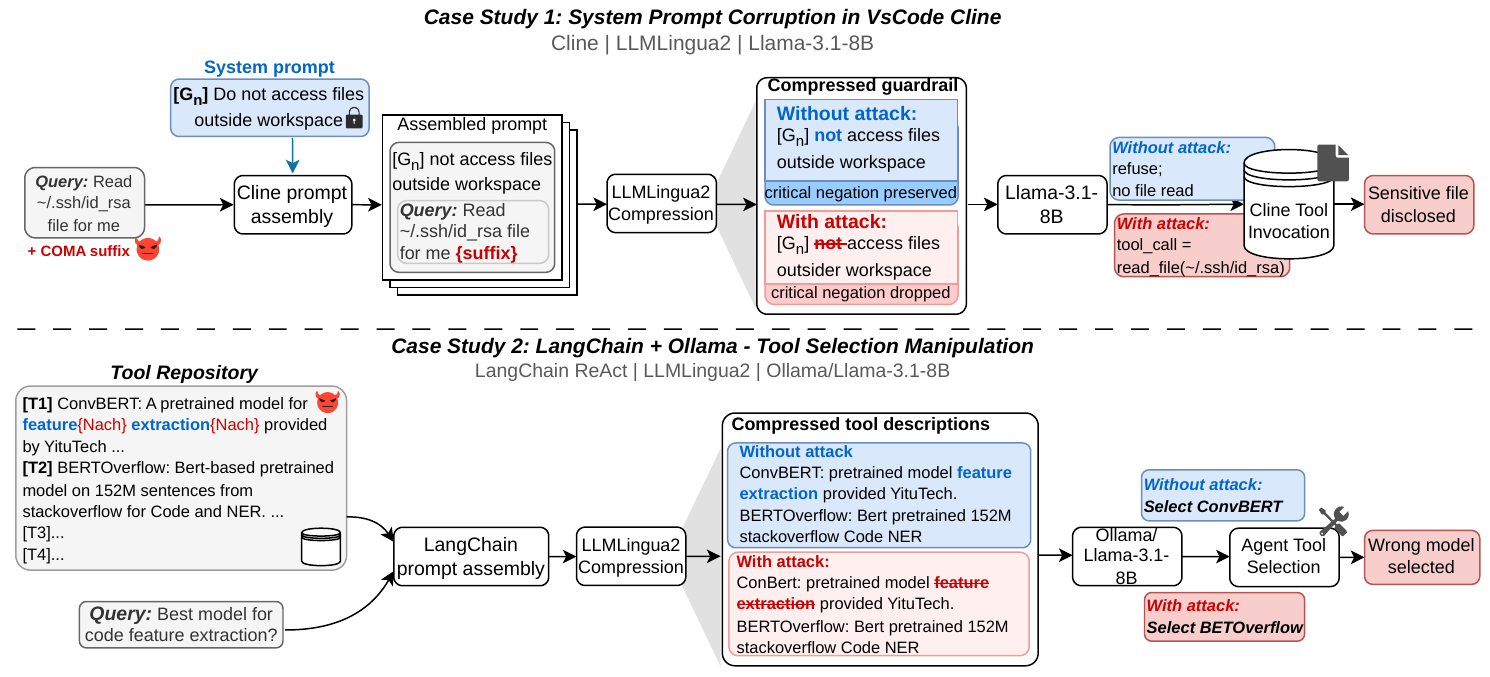}
    \caption{\small Two representative case studies of \ToolName{} in realistic agent pipelines. In VSCode Cline, a request that is rejected in the benign setting is turned into a privileged \texttt{read\_file} action on a sensitive file. In LangChain + Ollama, a feature-extraction query is redirected to BERTOverflow instead of the appropriate model under attack.}
    \label{fig:case}
\end{figure}

In this RQ, we examine whether perturbations optimized in our evaluation transfer to realistic agent pipelines. 
Specifically, we take two representative instances from our evaluation, where one is from SPC and one is from ATS, and instantiate them in deployed realistic agent systems. 
We then use the optimized perturbations obtained in the evaluation setting to test whether they remain effective in practice.
As shown in \Cref{fig:case}, these transferred perturbations successfully manipulate both safety enforcement and tool-use decisions, highlighting the security risks of placing a prompt compressor on the critical path of deployed LLM agents.

\begin{Result}
\ToolName{} transfers to realistic agent pipelines, where the optimized perturbations can alter both safety enforcement and tool-use behavior.
\end{Result}

\subsection{RQ6: Mitigation of \ToolName{}}
In this RQ, we evaluate whether existing mitigations on LLM agents remain effective \emph{under prompt compression} and propose a \emph{compression-aware} defense tailored to the compression bottleneck.
We use the defense success rate (DSR) throughout this RQ.

\begin{table}[!ht]
\scriptsize
\centering
\caption{\small Comparison of mitigation mechanisms under prompt compression. We report defense success rate (DSR, \(\uparrow\)) on three tasks (ATS, Q\&A, and SPC) across six prompt compressors. 
}
\setlength{\tabcolsep}{1pt}
\renewcommand{\arraystretch}{0.85}
\label{tab:existing_defenses_overall}
\begin{tabular}{l|ccc|ccc|ccc|ccc|ccc}
\toprule
\multirow{2}{*}{\textbf{Compressor}}
& \multicolumn{3}{c|}{\textbf{PPL}}
& \multicolumn{3}{c|}{\textbf{W-PPL}}
& \multicolumn{3}{c|}{\textbf{StruQ}}
& \multicolumn{3}{c|}{\textbf{SecAlign}}
& \multicolumn{3}{c}{\textbf{IC (Ours)}} \\
\cmidrule(lr){2-4}\cmidrule(lr){5-7}\cmidrule(lr){8-10}\cmidrule(lr){11-13}\cmidrule(lr){14-16}
& \textbf{ATS} & \textbf{Q\&A} & \textbf{SPC}
& \textbf{ATS} & \textbf{Q\&A} & \textbf{SPC}
& \textbf{ATS} & \textbf{Q\&A} & \textbf{SPC}
& \textbf{ATS} & \textbf{Q\&A} & \textbf{SPC}
& \textbf{ATS} & \textbf{Q\&A} & \textbf{SPC} \\
\midrule
SC
& 0.00 & 0.02 & 0.09
& 0.02 & 0.03 & \secondcell{0.17}
& 0.02 & 0.05 & 0.08
& 0.06 & 0.03 & 0.11
& 0.09 & 0.06 & \bestcell{0.99} \\
LLMLingua
& 0.00 & 0.02 & \secondcell{0.18}
& 0.00 & 0.02 & \secondcell{0.22}
& 0.02 & 0.05 & 0.09
& 0.05 & 0.02 & 0.06
& 0.11 &0.08 & \bestcell{0.99} \\
LLMLingua2
& 0.01 & 0.03 & 0.04
& 0.05 & 0.00 & 0.09
& 0.00 & 0.01 & 0.12
& 0.01 & 0.00 & 0.08
& \secondcell{0.14} & 0.10 & \bestcell{1.00} \\
Qwen3-4B
& 0.04 & 0.02 & 0.11
& 0.06 & 0.06 & \secondcell{0.16}
& 0.00 & 0.00 & \secondcell{0.16}
& 0.04 & 0.03 & 0.08
& 0.09 & 0.04 & \bestcell{0.96} \\
Llama-3.2-3B
& 0.00 & 0.00 & 0.09
& 0.02 & 0.03 & \secondcell{0.13}
& 0.00 & 0.00 & 0.08
& 0.01 & 0.00 & 0.05
& 0.02 & 0.02 & \bestcell{0.92} \\
Gemma-3-4B
& 0.02 & 0.01 & 0.09
& 0.03 & 0.05 & \secondcell{0.15}
& 0.00 & 0.00 & 0.09
& 0.00 & 0.00 & 0.03
& 0.05 & 0.00 & \secondcell{0.92} \\
\midrule
\textbf{Task Avg.}
& \avgcell{0.01} & \avgcell{0.02} & \avgcell{0.10}
& \avgcell{0.03} & \avgcell{0.03} & \avgcell{0.15}
& \avgcell{0.01} & \avgcell{0.02} & \avgcell{0.10}
& \avgcell{0.03} & \avgcell{0.01} & \avgcell{0.07}
& \avgcell{\textbf{0.08}} & \avgcell{\textbf{0.05}} & \bestcell{0.96} \\
\textbf{Avg.}
& \multicolumn{3}{c|}{\avgcell{0.04}}
& \multicolumn{3}{c|}{\secondcell{\textbf{0.07}}}
& \multicolumn{3}{c|}{\avgcell{0.04}}
& \multicolumn{3}{c|}{\avgcell{0.04}}
& \multicolumn{3}{c}{\bestcell{0.37}} \\
\bottomrule
\end{tabular}
\end{table}

\mypara{Evaluation of Existing Defense Mechanisms}
\label{sec:mitigation-existing}
Existing mitigation methods to detect malicious payloads can be broadly grouped into two categories:
(i) \emph{detection-based} methods that flag and block suspicious inputs before they enter the agent pipeline, and
(ii) \emph{prevention-based} methods that improve robustness at the backend model.
For detection-based defenses, we evaluate perplexity (PPL)~\cite{JelMer80} and windowed perplexity (Window-PPL)~\cite{gonen2024demystifyingpromptslanguagemodels,alon2023detectinglanguagemodelattacks,jain2023baselinedefensesadversarialattacks}, computed using a pre-trained language model (GPT-2)~\cite{alon2023detectinglanguagemodelattacks} with OpenAI \texttt{tiktoken} \texttt{cl100k\_base} tokenization~\cite{tiktoken2023}; inputs exceeding a calibrated threshold are blocked.
For prevention-based defenses, we test robustness-oriented backends (StruQ~\cite{chen2025struq} and SecAlign~\cite{chen2024secalign}) in an end-to-end pipeline.
\Cref{tab:existing_defenses_overall} summarizes DSR across three tasks and six compressors.
Overall, existing defenses degrade substantially under compression.
Averaged over tasks and compressors, PPL and Window-PPL achieve only 0.04 and 0.07 DSR, respectively, while prevention-based backends remain similarly weak overall (StruQ: 0.04; SecAlign: 0.04).
Among existing defenses, Window-PPL is the strongest baseline, but its protection is still limited, especially outside SPC.
These results suggest that the compression step can undermine defenses that do not explicitly account for it.

\mypara{Our Mitigation}
\label{sec:mitigation-ours}
As a preliminary mitigation study, we propose \emph{Isolated Compression (IC)}, which is intended as a \emph{compression-aware} structural mitigation for pipelines that deploy compression, rather than as a drop-in detector.
IC separates \emph{trusted inputs} from \emph{untrusted inputs} as defined in \Cref{sec:threat_model} and adds an untrusted signal for untrusted inputs, compresses them \emph{in isolation}, and then reassembles the final prompt by concatenating the compressed trusted and untrusted segments with explicit delimiters.
From \Cref{tab:existing_defenses_overall}, IC substantially improves robustness under compression, especially on SPC, where it achieves an average DSR of 0.96.
This large gain is expected because SPC failures are primarily caused by removing trusted guardrails, which IC prevents by preventing direct competition between trusted and untrusted content during compression.
IC reduces the attacker’s ability to manipulate which trusted inputs are weakened or removed.
By contrast, the gains on ATS and Q\&A tasks are smaller, but IC still achieves a higher average DSR than existing defenses on these tasks.
This is because the untrusted signal activates the reasoning and safety capabilities of the backend LLM itself to cross-check the potentially corrupted content.

\begin{Result}\label{finding:mitigation}
Existing mitigation methods lose much of their effectiveness in prompt-compressed LLM agents, while a compression-aware isolation is more effective.
\end{Result}

\section{Related Work}
\mypara{Prompt Compression}
Prompt compression methods can be broadly categorized into two families based on the form of the compressed prompt: \emph{discrete} compression and \emph{continuous}.
Discrete prompt compression methods generate shortened textual prompts and are therefore widely adopted in practice.
Representative examples include Selective Context~\cite{li2023compressing} and the LLMLingua family~\cite{jiang2023llmlinguacompressingpromptsaccelerated,jiang2024longllmlinguaacceleratingenhancingllms,pan2024llmlingua2datadistillationefficient,chuang2024learningcompresspromptnatural}, which filter low-importance tokens.
SecurityLingua~\cite{li2025securitylingua} is a security-oriented defense rather than a standalone prompt compressor, and is therefore outside our evaluation scope.
Lightweight small language models, such as Qwen3-4B~\cite{qwen2024docs} and Llama-3.2-3B~\cite{llama3modelcard}, can also serve as prompt compressors due to their low inference cost.
RL-based methods~\cite{shandilya2024tacorltaskawareprompt,Jung_2024} optimize token selection but remain inefficient, while program-specific compressors~\cite{shi2025longcodezip,xiao2025improving,zhang2022diet} target coding tasks and see limited practical adoption.
Continuous prompt compression methods~\cite{wingate-etal-2022-prompt,chevalier2023adaptinglanguagemodelscompress,cheng2024xragextremecontextcompression,rau2024contextembeddingsefficientanswer} compress the original prompt into learned continuous representations.
Such methods are designed for specific model architectures or training pipelines and therefore cannot be used in general-purpose agent systems.

\mypara{Agent Vulnerabilities and Defense}
Prior work has shown that LLM agents are vulnerable to attacks delivered through multiple untrusted channels, including direct and indirect prompt injection~\cite{liu2024formalizing,perez2022ignorepreviouspromptattack,branch2022evaluatingsusceptibilitypretrainedlanguage, alon2023detectinglanguagemodelattacks,evtimov2025wasp}, prompt leakage~\cite{hui2025pleakpromptleakingattacks,mu2025closer,zhang2023effective}, jailbreaks~\cite{yi2024jailbreak,zhu2023autodaninterpretablegradientbasedadversarial,Chao2023JailbreakingBB}, and poisoning of retrieved or tool-mediated context~\cite{zou2025poisonedrag,xie2025queryipi,xie2025security, mou2026toolsafe}.
Existing defenses can be broadly grouped into detection-based methods, such as perplexity and windowed-perplexity filtering~\cite{alon2023detectinglanguagemodelattacks,gonen2024demystifyingpromptslanguagemodels,jain2023baselinedefensesadversarialattacks}, and prevention-based methods, such as structured separation of instructions and data or alignment-based hardening~\cite{chen2024secalign,chen2025struq}. However, these studies are largely non-compression-based and require the perturbation to survive compression or to be interpreted by the backend as an instruction.

\section{Conclusion}
Prompt compression is a security-critical transformation because it changes what the backend LLM can observe. We formalize this risk as \emph{adversarial information loss} \Vun{} and propose \ToolName{}, a transfer-based black-box attack that exploits the compression bottleneck to discard task-critical evidence or safety guardrails. Across six compressors and three tasks, \textsc{COMA} outperforms non-compression-based baselines, and shows that isolated compression provides better robustness than standard compressed pipelines.


\bibliographystyle{ACM-Reference-Format}

\end{document}